\documentclass[11pt,onecolumn,aps,superscriptaddress,
               amsmath,amssymb,nofootinbib]{revtex4}      
\usepackage{graphicx}

\usepackage{graphicx,epsfig}

\usepackage{dcolumn}
\usepackage{bm}

\newcommand{\Z}{{\mathbb Z}}

\begin{document}

\begin{flushright}
\baselineskip=12pt \normalsize
{ACT-08-09},
{MIFP 09-36}\\
\smallskip
\end{flushright}

\title{The Search for a Realistic String Model at LHC}

\author{James A. Maxin}
\affiliation{George P. and Cynthia W. Mitchell Institute for
Fundamental Physics, Texas A\&M
University,\\ College Station, TX 77843, USA}
\author{Van E. Mayes}
\affiliation{George P. and Cynthia W. Mitchell Institute for
Fundamental Physics, Texas A\&M
University,\\ College Station, TX 77843, USA}
\author{D.V. Nanopoulos}
\affiliation{George P. and Cynthia W. Mitchell Institute for
Fundamental Physics, Texas A\&M
University,\\ College Station, TX 77843, USA}
\affiliation{Astroparticle Physics Group, Houston
Advanced Research Center (HARC),
Mitchell Campus,
Woodlands, TX~77381, USA; \\
Academy of Athens,
Division of Natural Sciences, 28~Panepistimiou Avenue, Athens 10679,
Greece}

\begin{abstract}
\begin{center}
{\bf ABSTRACT}
\end{center}

We survey the low-energy supersymmetry phenomenology of a three-family Pati-Salam model constructed from intersecting $D$6-branes in
Type IIA string theory on the $\mathbf{T^6/(\Z_2\times \Z_2)}$ orientifold which possesses many of the phenomenological properties desired in string model-building. In the model, there is no exotic matter in the low-energy spectrum, the correct mass hierarchies for quarks and leptons may be obtained, and the gauge couplings are automatically unified at the string scale. We calculate the supersymmetry breaking soft terms and the corresponding low-energy supersymmetry particle spectra for the model. We find the WMAP constrained dark matter density can be generated in this model in the stau-neutralino and chargino-neutralino coannihilation regions, with expected final states at LHC consisting of low energy leptons and $\cal{O}$(GeV) neutrinos. Moreover, we expect final states in the supercritical string cosmology (SSC) scenario to comprise high energy leptons and $\cal{O}$(GeV) neutrinos.
\end{abstract}

\maketitle

\newpage
\section{Introduction}

The main goal of string phenomenology is to make contact between 
string theory and the real world.  In particular, this involves searching for a specific string vacuum 
which reproduces the Standard Model (SM) in complete detail.  This is not an easy task. 
The SM 
has an intricate structure, with
three-generations of chiral fermions which transform as
bifundamental representations of $SU(3)_C \times SU(2)_L \times
U(1)_Y$. In addition to the fact that the SM fermions are
replicated into three distinct families, the families
exhibit a pattern of mass hierarchies and
mixings.  Although there have been many models which can reproduce the
gross features of the SM, there are generally problems either with 
extra exotic particles or an inability to generate fermion mass hierarchies
and mixings.  

Interestingly, the so-called intersecting $D$6-brane models
where the chiral
fermions arise at the intersections between $D$6 branes
(Type IIA)
in the internal space~\cite{bdl} together with the T-dual Type IIB
description in terms of magnetized D-branes~\cite{bachas} have
provided an exciting approach toward constructing semi-realistic
string vacua (for reviews, see~\cite{Blumenhagen:2005mu,
Blumenhagen:2006ci}).
In particular, intersecting D-brane models may naturally
generate the SM fermion mass hierarchies and mixings, as well as
an explanation for the replication of chirality.
Indeed, such models provide promising setups
which may accommodate semi-realistic features of low-energy
physics.  
In short,
$D$6-branes (in Type IIA string theory) fill four-dimensional Minkowski space-time
and wrap 3-cycles in the compact manifold, with a stack of $N$
$D$6-branes having a gauge group $U(N)$ (or $U(N/2)$ in the case of
$\mathbf{T^6/(\Z_2 \times \Z_2)}$) in its world volume. The
3-cycles wrapped by the D-branes will in general intersect
multiple times in the internal space, resulting in chiral fermions
in the bifundamental representation localized at the intersections
between different stacks. The multiplicity of such fermions is
then given by the number of times the 3-cycles intersect.

The Yukawa couplings in intersecting $D$6-brane models arise from
open string world-sheet instantons that connect three $D$6-brane
intersections \cite{Aldazabal:2000cn}. For a given triplet of
intersections, the minimal world-sheet action which contributes to
the trilinear Yukawa couplings is weighted by a factor
$\exp(-A_{abc})$, where $A_{abc}$ is the world-sheet area of the
triangle bounded by the branes $a, b,$ and $c$. Since there are
several possible triangles with different areas, mass hierarchies
may inherently arise. The Yukawa couplings depend on both the
D-brane positions in the internal space as well as on the geometry
of the underlying compact manifold. Effectively, these quantities
are parameterized by the vacuum expectation values (VEVs) of open
and closed-string moduli.

In most intersecting D-brane models, there has typically been a rank
one problem in the SM fermion Yukawa matrices, preventing the
generation of masses and mixings for the first two families of
quarks and leptons. For the case of toroidal orientifold
compactifications, this can be traced to the fact that not all of
the SM fermions are localized at intersections on the same
torus~\cite{Cremades:2003qj, Chamoun:2003pf, Kitazawa:2004nf,
Dutta:2005bb}. However, there is one example known of an intersecting $D$6-brane
model in Type IIA on the $\mathbf{T^6/(\Z_2\times \Z_2)}$
orientifold where these problems
may be solved~\cite{Cvetic:2004ui,Chen:2006gd}.  Thus, this
particular model may be a step forward to obtaining realistic
phenomenology from string theory. Indeed, as we have recently
shown~\cite{Chen:2007px, Chen:2007zu}, it is possible within the moduli space
of this model to obtain the correct SM quark masses and mixings,
the tau lepton mass, and to generate naturally small neutrino
masses via the seesaw mechanism. In addition to these features,
the model exhibits automatic gauge coupling unification, and it is
possible to generate realistic low-energy supersymmetric particle
spectra, a subset of which may produce the observed dark matter
density~\cite{Chen:2007px, Chen:2007zu}.

Although this model exhibits a realistic chiral sector, it cannot be considered
fully realistic until the moduli stabilization problem has been completely addressed.
For example, although it has been shown that it is possible to obtain correct 
Yukawa mass matrices, it is not possible to say that this is a unique solution
until both open and closed string moduli VEVs can be fixed dynamically.  Similar considerations
apply to the case of supersymmetry breaking, although the low-energy effective action
may be determined as functions of the moduli.  However, in light of the soon to be operational 
Large Hadron Collider (LHC), it is still an interesting exercise to study the possible phenomenology of this model which could potentially
be observed at LHC, which is the subject of this paper.

\section{The Model}

In this section, we briefly describe the intersecting D-brane model under study. 
We consider Type IIA string theory
compactified on a $\mathbf{T^6/(\Z_2\times \Z_2)}$
orientifold~\cite{CSU}. The $\mathbf{T^{6}}$ is a six-torus
factorized as $\mathbf{T^{6}} = \mathbf{T^2} \times \mathbf{T^2}
\times \mathbf{T^2}$ whose complex coordinates are $z_i$, $i=1,\;
2,\; 3$ for the $i^{th}$ two torus, respectively. The $\theta$ and
$\omega$ generators for the orbifold group $\mathbf{\Z_{2} \times \Z_{2}}$,
act on the complex coordinates of $\mathbf{T^6}$ as
\begin{eqnarray}
& \theta: & (z_1,z_2,z_3) \to (-z_1,-z_2,z_3)~,~ \nonumber \\
& \omega: & (z_1,z_2,z_3) \to (z_1,-z_2,-z_3)~.~\,
\label{orbifold}
\end{eqnarray}
The orientifold projection is applied by gauging the symmetry
$\Omega R$, where $\Omega$ is world-sheet parity, and $R$ is given
by
\begin{eqnarray}
R: (z_1,z_2,z_3) \to ({\overline z}_1,{\overline z}_2,{\overline
z}_3)~.~\,    \label{orientifold}
\end{eqnarray}
Thus, there are four kinds of orientifold 6-planes (O6-planes) for
the actions $\Omega R$, $\Omega R\theta$, $\Omega R \omega$, and
$\Omega R\theta\omega$, respectively. There are two kinds of
complex structures consistent with orientifold projection for a
two torus: rectangular and tilted~\cite{CSU}. If we denote the
homology classes of the three cycles wrapped by the $D$6-brane
stacks as $n_P^i[a_i]+m_P^i[b_i]$ and $n_P^i[a'_i]+m_P^i[b_i]$
with $[a_i']=[a_i]+\frac{1}{2}[b_i]$ for the rectangular and
tilted tori respectively, we can label a generic one cycle by
$(n_P^i,l_P^i)$ in either case, where in terms of the wrapping
numbers $l_{P}^{i}\equiv m_{P}^{i}$ for a rectangular two torus
and $l_{P}^{i}\equiv 2\tilde{m}_{P}^{i}=2m_{P}^{i}+n_{P}^{i}$ for
a tilted two torus. Moreover, for a stack of $N$ $D$6-branes that
does not lie on one of the O6-planes, we obtain a $U(N/2)$ gauge
symmetry with three adjoint chiral superfields due to the orbifold
projections, while for a stack of $N$ $D$6-branes which lies on an
O6-plane, we obtain a $USp(N)$ gauge symmetry with three
anti-symmetric chiral superfields. Bifundamental chiral
superfields arise from the intersections of two different stacks
$P$ and $Q$  of $D$6-branes or from one stack $P$ and its $\Omega R$
image $P'$~\cite{CSU}.

We present the $D$6-brane configurations and intersection numbers of
the model in Table~\ref{MI-Numbers}, and the resulting spectrum in
Table~\ref{Spectrum}~\cite{Cvetic:2004ui,Chen:2006gd}. We put the $a'$, $b$,
and $c$ stacks of $D$6-branes on top of each other on the third
two torus, and as a result there are additional vector-like
particles from $N=2$ subsectors.

\begin{table}[htb]
\footnotesize
\renewcommand{\arraystretch}{1.0}
\caption{$D$6-brane configurations and intersection numbers.}
\label{MI-Numbers}
\begin{center}
\begin{tabular}{|c||c|c||c|c|c|c|c|c|c|c|c|c|}
\hline
& \multicolumn{12}{c|}{$U(4)_C\times U(2)_L\times U(2)_R\times USp(2)^4$}\\
\hline \hline  & $N$ & $(n^1,l^1)\times (n^2,l^2)\times

(n^3,l^3)$ & $n_{S}$& $n_{A}$ & $b$ & $b'$ & $c$ & $c'$& 1 & 2 & 3 & 4 \\

\hline

    $a$&  8& $(0,-1)\times (1,1)\times (1,1)$ & 0 & 0  & 3 & 0 & -3 & 0 & 1 & -1 & 0 & 0\\

    $b$&  4& $(3,1)\times (1,0)\times (1,-1)$ & 2 & -2  & - & - & 0 & 0 & 0 & 1 & 0 & -3 \\

    $c$&  4& $(3,-1)\times (0,1)\times (1,-1)$ & -2 & 2  & - & - & - & - & -1 & 0 & 3 & 0\\

\hline

    1&   2& $(1,0)\times (1,0)\times (2,0)$ & \multicolumn{10}{c|}{$\chi_1=3,~
\chi_2=1,~\chi_3=2$}\\

    2&   2& $(1,0)\times (0,-1)\times (0,2)$ & \multicolumn{10}{c|}{$\beta^g_1=-3,~
\beta^g_2=-3$}\\

    3&   2& $(0,-1)\times (1,0)\times (0,2)$& \multicolumn{10}{c|}{$\beta^g_3=-3,~
\beta^g_4=-3$}\\

    4&   2& $(0,-1)\times (0,1)\times (2,0)$ & \multicolumn{10}{c|}{}\\

\hline

\end{tabular}

\end{center}

\end{table}

\begin{table}[htb]

\footnotesize

\renewcommand{\arraystretch}{1.0}

\caption{The chiral and vector-like superfields, and their quantum
numbers under the gauge symmetry $SU(4)_C\times SU(2)_L\times
SU(2)_R \times USp(2)_1 \times USp(2)_2 \times USp(2)_3 \times
USp(2)_4$.}

\label{Spectrum}

\begin{center}

\begin{tabular}{|c||c||c|c|c||c|c|c|}\hline

 & Quantum Number

& $Q_4$ & $Q_{2L}$ & $Q_{2R}$  & Field \\

\hline\hline

$ab$ & $3 \times (4,\overline{2},1,1,1,1,1)$ & 1 & -1 & 0  & $F_L(Q_L, L_L)$\\

$ac$ & $3\times (\overline{4},1,2,1,1,1,1)$ & -1 & 0 & $1$   & $F_R(Q_R, L_R)$\\

$a1$ & $1\times (4,1,1,2,1,1,1)$ & $1$ & 0 & 0  & \\

$a2$ & $1\times (\overline{4},1,1,1,2,1,1)$ & -1 & 0 & 0   & \\

$b2$ & $1\times(1,2,1,1,2,1,1)$ & 0 & 1 & 0    & \\

$b4$ & $3\times(1,\overline{2},1,1,1,1,2)$ & 0 & -1 & 0    & \\

$c1$ & $1\times(1,1,\overline{2},2,1,1,1)$ & 0 & 0 & -1    & \\

$c3$ & $3\times(1,1,2,1,1,2,1)$ & 0 & 0 & 1   &  \\

$b_{S}$ & $2\times(1,3,1,1,1,1,1)$ & 0 & 2 & 0   &  $T_L^i$ \\

$b_{A}$ & $2\times(1,\overline{1},1,1,1,1,1)$ & 0 & -2 & 0   & $S_L^i$ \\

$c_{S}$ & $2\times(1,1,\overline{3},1,1,1,1)$ & 0 & 0 & -2   & $T_R^i$  \\

$c_{A}$ & $2\times(1,1,1,1,1,1,1)$ & 0 & 0 & 2   & $S_R^i$ \\

\hline\hline

$ab'$ & $3 \times (4,2,1,1,1,1,1)$ & 1 & 1 & 0  & \\

& $3 \times (\overline{4},\overline{2},1,1,1,1,1)$ & -1 & -1 & 0  & \\

\hline

$ac'$ & $3 \times (4,1,2,1,1,1,1)$ & 1 &  & 1  & $\Phi_i$ \\

& $3 \times (\overline{4}, 1, \overline{2},1,1,1,1)$ & -1 & 0 & -1  &
$\overline{\Phi}_i$\\

\hline

$bc$ & $6 \times (1,2,\overline{2},1,1,1,1)$ & 0 & 1 & -1   & $H_u^i$, $H_d^i$\\

& $6 \times (1,\overline{2},2,1,1,1,1)$ & 0 & -1 & 1   & \\

\hline

\end{tabular}

\end{center}

\end{table}

The model resulting from this configuration is a three-family Pati-Salam 
model with gauge group $U(4)\times U(2)_L \times U(1)_R$. 
The anomalies from three global $U(1)$s of $U(4)_C$, $U(2)_L$ and
$U(2)_R$ are canceled by a generalized Green-Schwarz mechanism, which results in the
gauge fields of these $U(1)$s obtaining masses via the linear
$B\wedge F$ couplings. Thus, the effective gauge symmetry is
$SU(4)_C\times SU(2)_L\times SU(2)_R$. 

In order to break the gauge
symmetry to the SM, we split the $a$ stack of $D$6-branes
into $a_1$ and $a_2$ stacks, with $N_{a_1}=6$ and $N_{a_2}=2$ respectively,
and split the $c$ stack of $D$6-branes into $c_1$ and $c_2$ stacks
with $N_{c_1}=N_{c_2}=2$. 
In this way, the gauge symmetry
is further broken to $ SU(3)_C\times SU(2)_L\times
U(1)_{I_{3R}}\times U(1)_{B-L}$. Moreover, the
$U(1)_{I_{3R}}\times U(1)_{B-L}$ gauge symmetry may be broken to
$U(1)_Y$ by giving vacuum expectation values (VEVs) to the
vector-like particles with the quantum numbers $({\bf { 1}, 1,
1/2, -1})$ and $({\bf { 1}, 1, -1/2, 1})$ under the $SU(3)_C\times
SU(2)_L\times U(1)_{I_{3R}} \times U(1)_{B-L} $ gauge symmetry
from $a_2 c_1'$ intersections~\cite{Cvetic:2004ui,Chen:2006gd}.  Thus, we
obtain a three-family Standard Model preserving $\mathcal{N}=1$ 
supersymmetry. 

Using the values for the complex structure moduli
obtained from the conditions for preserving $\mathcal{N}=1$ supersymmetry, it has been found that 
the SM gauge couplings are automatically unified at the string scale~\cite{Chen:2007px, Chen:2007zu}.  In addition,
after fixing the unified value of the gauge coupling constant
at the string scale to that obtained in the MSSM, the hidden
sector gauge groups become confining at high mass scales, thus matter charged under these groups
is decoupled~\cite{Chen:2007px, Chen:2007zu}.  
Therefore, this model exhibits a completely realistic chiral sector.  Although at this 
point it is not possible to make definitive predictions until the moduli stabilization problem has
been adequately addressed, it 
is of some interest to study the low-energy supersymmetric phenomenology of this model in regards
to potential observations at LHC.  

\section{The $\mathcal{N}=1$ Low-energy Effective Action}

To discuss the low-energy phenomenology, we start from the 
$\mathcal{N}=1$ low-energy
effective action. From the effective scalar potential it is
possible to study the stability~\cite{Blumenhagen:2001te}, the
tree-level gauge couplings \cite{CLS1, Shiu:1998pa,
Cremades:2002te}, gauge threshold corrections \cite{Lust:2003ky},
and gauge coupling unification \cite{Antoniadis:Blumen}.  The
effective Yukawa couplings \cite{Cremades:2003qj, Cvetic:2003ch},
matter field K\"ahler metric and soft-SUSY breaking terms have
also been investigated \cite{Kors:2003wf}.  A more detailed
discussion of the K\"ahler metric and string scattering of gauge,
matter, and moduli fields has been performed in
\cite{Lust:2004cx} (Also see refs~\cite{Misra:2007cq,Misra:2008tx,Misra:2009ei,Bertolini:2005qh,Billo:2007py}). In principle, it should be possible to
specify the exact mechanism by which supersymmetry is broken once the
moduli stabilization problem has been solved, and
thus to make very specific predictions.  However, for the present
work, we will adopt a parametrization of the SUSY breaking so that
we can study it generically.

The $\mathcal{N}=1$ supergravity action depends upon three
functions, the holomorphic gauge kinetic function $f$, K\a"ahler
potential $K$, and the superpotential $W$.  Each of these will in
turn depend upon the moduli fields which describe the background
upon which the model is constructed. The holomorphic gauge kinetic
function for a $D$6-brane wrapping a calibrated three-cycle is given
by~\cite{Blumenhagen:2006ci}
\begin{equation}
f_P = \frac{1}{2\pi \ell_s^3}\left[e^{-\phi}\int_{\Pi_P} \mbox{Re}(e^{-i\theta_P}\Omega_3)-i\int_{\Pi_P}C_3\right].
\end{equation}
In terms of the three-cycle wrapped by the stack of branes, we have
\begin{equation}
\int_{\Pi_a}\Omega_3 = \frac{1}{4}\prod_{i=1}^3(n_a^iR_1^i + 2^{-\beta_i}il_a^iR_2^i).
\end{equation}
from which it follows that
\begin{eqnarray}
f_P &=&
\frac{1}{4\kappa_P}(n_P^1\,n_P^2\,n_P^3\,s-\frac{n_P^1\,l_P^2\,l_P^3\,u^1}{2^{(\beta_2+\beta_3)}}-\frac{n_P^2\,l_P^1\,l_P^3\,u^2}{2^{(\beta_1+\beta_3)}}-
\frac{n_P^3\,l_P^1\,l_P^2\,u^3}{2^{(\beta_1+\beta_2)}}),
\label{kingauagefun}
\end{eqnarray}
where $\kappa_P = 1$ for $SU(N_P)$ and $\kappa_P = 2$ for
$USp(2N_P)$ or $SO(2N_P)$ gauge groups and where we use the $s$ and
$u$ moduli in the supergravity basis.  In the string theory basis,
we have the dilaton $S$, three K\"ahler moduli $T^i$, and three
complex structure moduli $U^i$~\cite{Lust:2004cx}. These are related to the
corresponding moduli in the supergravity basis by
\begin{eqnarray}
\mathrm{Re}\,(s)& =&
\frac{e^{-{\phi}_4}}{2\pi}\,\left(\frac{\sqrt{\mathrm{Im}\,U^{1}\,
\mathrm{Im}\,U^{2}\,\mathrm{Im}\,U^3}}{|U^1U^2U^3|}\right)
\nonumber \\
\mathrm{Re}\,(u^j)& =&
\frac{e^{-{\phi}_4}}{2\pi}\left(\sqrt{\frac{\mathrm{Im}\,U^{j}}
{\mathrm{Im}\,U^{k}\,\mathrm{Im}\,U^l}}\right)\;
\left|\frac{U^k\,U^l}{U^j}\right| \qquad (j,k,l)=(\overline{1,2,3})
\nonumber \\
\mathrm{Re}(t^j)&=&\frac{i\alpha'}{T^j} \label{idb:eq:moduli}
\end{eqnarray}
and $\phi_4$ is the four-dimensional dilaton.
To second order in the string matter fields, the K\a"ahler potential is given by
\begin{eqnarray}
K(M,\bar{M},C,\bar{C}) = \hat{K}(M,\bar{M}) + \sum_{\mbox{untwisted}~i,j} \tilde{K}_{C_i \bar{C}_j}(M,\bar{M})C_i \bar{C}_j + \\ \nonumber \sum_{\mbox{twisted}~\theta} \tilde{K}_{C_{\theta} \bar{C}_{\theta}}(M,\bar{M})C_{\theta}\bar{C}_\theta.
\end{eqnarray}
The untwisted moduli $C_i$, $\bar{C}_j$ are light, non-chiral
scalars from the field theory point of view, associated with the
D-brane positions and Wilson lines.  These fields are not observed
in the MSSM, and if they were present in the low energy spectra
may disrupt the gauge coupling unification.  Clearly, these
fields must get a large mass through some mechanism, and for the present 
it is assumed that the open-string moduli become massive via high-dimensional
operators.

For twisted moduli arising from strings stretching between stacks
$P$ and $Q$, we have $\sum_j\theta^j_{PQ}=0$, where $\theta^j_{PQ} =
\theta^j_Q - \theta^j_P$ is the angle between the cycles wrapped
by the stacks of branes $P$ and $Q$ on the $j^{th}$ torus
respectively. Then, for the K\a"ahler metric in Type IIA theory we find
the following two cases:

\begin{itemize}

\item $\theta^j_{PQ}<0$, $\theta^k_{PQ}>0$, $\theta^l_{PQ}>0$

\begin{eqnarray}
\tilde{K}_{PQ} &=& e^{\phi_4} e^{\gamma_E (2-\sum_{j = 1}^3
\theta^j_{PQ}) }
\sqrt{\frac{\Gamma(\theta^j_{PQ})}{\Gamma(1+\theta^j_{PQ})}}
\sqrt{\frac{\Gamma(1-\theta^k_{PQ})}{\Gamma(\theta^k_{PQ})}}
\sqrt{\frac{\Gamma(1-\theta^l_{PQ})}{\Gamma(\theta^l_{PQ})}}
\nonumber \\ && (t^j + \bar{t}^j)^{\theta^j_{PQ}} (t^k +
\bar{t}^k)^{-1+\theta^k_{PQ}} (t^l +
\bar{t}^l)^{-1+\theta^l_{PQ}}.
\end{eqnarray}

\item $\theta^j_{PQ}<0$, $\theta^k_{PQ}<0$, $\theta^l_{PQ}>0$

\begin{eqnarray}
\tilde{K}_{PQ} &=& e^{\phi_4} e^{\gamma_E (2+\sum_{j = 1}^3
\theta^j_{PQ}) }
\sqrt{\frac{\Gamma(1+\theta^j_{PQ})}{\Gamma(-\theta^j_{PQ})}}
\sqrt{\frac{\Gamma(1+\theta^k_{PQ})}{\Gamma(-\theta^k_{PQ})}}
\sqrt{\frac{\Gamma(\theta^l_{PQ})}{\Gamma(1-\theta^l_{PQ})}}
\nonumber \\ && (t^j + \bar{t}^j)^{-1-\theta^j_{PQ}} (t^k +
\bar{t}^k)^{-1-\theta^k_{PQ}} (t^l + \bar{t}^l)^{-\theta^l_{PQ}}.
\end{eqnarray}

\end{itemize}

For branes which are parallel on at least one torus, giving rise
to non-chiral matter in bifundamental representations (for example,
the Higgs doublets), the K\a"ahler metric is
\begin{equation}
\hat{K}=((s+\bar{s})(t^1+\bar{t}^1)(t^2+\bar{t}^2)(u^3+\bar{u}^3))^{-1/2}.
\label{nonchiralK}
\end{equation}
The superpotential is given by
\begin{equation}
W = \hat{W}+ \frac{1}{2}\mu_{\alpha\beta}(M)C^{\alpha}C^{\beta} + \frac{1}{6}Y_{\alpha\beta\gamma}(M)C^{\alpha\beta\gamma}+\cdots
\end{equation}
while the minimum of the F part of the tree-level supergravity
scalar potential $V$ is given by
\begin{equation}
V(M,\bar{M}) = e^G(G_M K^{MN} G_N -3) = (F^N K_{NM} F^M-3e^G),
\end{equation}
where
$G_M=\partial_M G$ and $K_{NM}=\partial_N \partial_M K$, $K^{MN}$
is inverse of $K_{NM}$, and the auxiliary fields $F^M$ are given
by
\begin{equation}
F^M=e^{G/2} K^{ML}G_L. \label{aux}
\end{equation}
Supersymmetry is broken when some of the F-terms of the hidden sector fields $M$
acquire VEVs. This then results in soft terms being generated in
the observable sector. For simplicity, it is assumed in this
analysis that the $D$-term does not contribute (see
\cite{Kawamura:1996ex}) to the SUSY breaking.  Then the goldstino
is included by the gravitino via the superHiggs effect. The
gravitino then obtains a mass
\begin{equation}
m_{3/2}=e^{G/2},
\end{equation} 
The normalized gaugino mass parameters, scalar mass-squared
parameters, and trilinear parameters respectively may be given in
terms of the K\a"ahler potential, the gauge kinetic function, and
the superpotential as
\begin{eqnarray}
M_P &=& \frac{1}{2\mbox{Re}f_P}(F^M\partial_M f_P), \\ \nonumber
m^2_{PQ} &=& (m^2_{3/2} + V_0) - \sum_{M,N}\bar{F}^{\bar{M}}F^N\partial_{\bar{M}}\partial_{N}log(\tilde{K}_{PQ}), \\ \nonumber
A_{PQR} &=& F^M\left[\hat{K}_M + \partial_M log(Y_{PQR}) - \partial_M log(\tilde{K}_{PQ}\tilde{K}_{QR}\tilde{K}_{RP})\right],
\label{softterms}
\end{eqnarray}
where $\hat{K}_M$ is the K\a"ahler metric appropriate for branes
which are parallel on at least one torus, i.e. involving
non-chiral matter.  

The above formulas for the soft terms depend on the Yukawa couplings, via the superpotential.  An important consideration is whether or not this should cause any modification to the low-energy spectrum.  However, this turns out not to be the case since the Yukawas in the soft term formulas are not the same as the physical Yukawas, which arise from world-sheet instantons and are proportional to $exp({-A})$, where $A$ is the world-sheet area of the triangles formed by a triplet of intersections at which
the Standard Model fields are localized.  As we shall see in a later section, the physical Yukawa couplings in Type IIA depend on the K\a"ahler moduli and the open-string moduli.  This ensures that the Yukawa couplings present in the soft terms do not depend on either the complex-structure moduli or dilaton (in the supergravity basis).  Thus, the
Yukawa couplings will not affect the low-energy spectrum in the case of $u$-moduli dominant and mixed $u$ and $s$ dominant supersymmetry breaking.

To determine the soft-supersymmetry breaking
parameters, and therefore the spectra of the models, we introduce
the VEVs of the auxiliary fields Eq. (\ref{aux}) for the
dilaton, complex and K\"ahler moduli \cite{Brignole:1993dj}:
\begin{eqnarray}
&& F^s=2\sqrt{3}C m_{3/2} {\rm Re}(s) \Theta_s e^{-i\gamma_s},
\nonumber \\
&&F^{\{u,t\}^i} = 2\sqrt{3}C m_{3/2}( {\rm Re}  ({u}^i) \Theta_i^u
e^{-i\gamma^u_i}+  {\rm Re} ({t}^i) \Theta_i^t
e^{-i\gamma_i^t}).
\end{eqnarray}
The factors $\gamma_s$ and $\gamma_i$ are the CP violating phases of the moduli, while
the constant $C$
is given by
\begin{equation}
C^2 = 1+ \frac{V_0}{3 m^2_{3/2}}.
\end{equation}
The goldstino is absorbed into the gravitino by $\Theta_S$ in $S$
field space, and $\Theta_i$ parameterize the goldstino direction
in $U^i$ space,  where $\sum (|\Theta_i^u|^2 + |\Theta_i^t|^2) + |\Theta_s|^2 =1$. The
goldstino angle $\Theta_s$ determines the degree to which SUSY
breaking is being dominated by the dilaton $s$ and/or complex
structure ($u^i$) and K\"ahler ($t^i$) moduli.  As suggested earlier, we
will not consider the case of $t$-moduli dominant supersymmetry breaking 
as in this case, the soft terms are not independent of the Yukawa couplings.

Next, we turn our attention to the soft-supersymmetry breaking
terms at the Grand Unification Theory (GUT) scale defined in Eq.~(\ref{softterms}).  In the
present analysis, not all the F-terms of the moduli get VEVs for simplicity, as
in \cite{Font:2004cx, Kane:2004hm}. As discussed earlier, we will assume that $F^t_i=0$ so that the soft terms have
no dependence on the physical Yukawa couplings. 

For the present work we will consider $u$-moduli dominated SUSY breaking where both the cosmological
constant $V_0$ and the goldstino angle are set to zero, such that $F^s=F^{t^i}=0$.   
Thus, we take $\Theta_s = 0$ so that the $F$-terms are parameterized by the expression
\begin{equation}
F^{u^i} = \sqrt{3}m_{3/2}(u^i + \bar{u}^i)\Theta_i e^{-i\gamma_i},
\label{auxfields}
\end{equation}
where $i = 1\mbox{,} 2\mbox{,} 3$ and with $\sum |\Theta_i|^2 =
1$. With this parametrization, the gaugino mass terms for a stack
$P$ may be written as
\begin{eqnarray}
M_P=\frac{-\sqrt{3}m_{3/2}}{\mbox{Re}
f_P}\sum_{j=1}^3 \left(\mbox{Re} u^j\,\Theta_j\,
e^{-i\gamma_j}\,n^j_Pm^k_Pm^l_P\right) \qquad
(j,k,l)=(\overline{1,2,3}).
\label{eq:idb:gaugino}
\end{eqnarray}
The Bino mass parameter is a linear combination of the
gaugino mass for each stack,
\begin{equation}
M_Y = \frac{1}{f_Y}\sum_P c_P M_P
\label{eq:idb:bino}
\end{equation}
where the the coefficients $c_P$ correspond to the linear
combination of $U(1)$ factors
which define the hypercharge, $U(1)_Y = \sum c_P U(1)_P$.

For the trilinear parameters, we have
\begin{eqnarray}
A_{PQR}&=&-\sqrt{3}m_{3/2}\sum_{j=1}^3 \left[
\Theta_je^{-i\gamma_j}\left(1+(\sum_{k=1}^3
 \xi_{PQ}^{k,j}\Psi(\theta^k_{PQ})-\frac{1}{4})+(\sum_{k=1}^3
 \xi_{RP}^{k,j}\Psi(\theta^k_{RP})-\frac{1}{4})
\right)\right]  \nonumber \\
&&+\frac{\sqrt{3}}{2}m_{3/2}{\Theta}_{3}e^{-i{\gamma}_1}
\label{eq:idb:trilinear_u}
\end{eqnarray}
where $P$,$Q$, and $R$ label the stacks of branes whose mutual
intersections define the fields present in the corresponding
trilinear coupling and the angle differences are defined as
\begin{equation}
\theta_{PQ} = \theta_Q - \theta_P.
\end{equation}
We must be careful when dealing with cases where the angle difference is
negative.  Note for the present model, there is always either one
or two of the $\theta_{PQ}$ which are negative.  Let us define the
parameter
\begin{equation}
\eta_{PQ} = \mbox{sgn}(\prod_i \theta_{PQ}^i),
\end{equation}
such that $\eta_{PQ} = -1$ indicates that only one of the angle
differences is negative while $\eta_{PQ} = +1$ indicates that two
of the angle differences are negative.

Finally, the squark and slepton (1/4 BPS) scalar mass-squared
parameters are given by
\begin{eqnarray}
m^2_{PQ}= m_{3/2}^2\left[1-3\sum_{m,n=1}^3
\Theta_m\Theta_ne^{-i(\gamma_m-\gamma_n)}\left(
\frac{{\delta}_{mn}}{4}+ \sum_{j=1}^3 (\xi^{j,m\bar
n}_{PQ}\Psi(\theta^j_{PQ})+
 \xi^{j,m}_{PQ}\xi^{j,\bar n}_{PQ}\Psi'(\theta^j_{PQ}))\right)
\right].
\label{eq:idb:scalarmass_u}
\end{eqnarray}

The functions $\Psi(\theta_{PQ})=\frac{\partial \ln
(e^{-\phi_4}\tilde{K}_{PQ})}{\partial \theta_{PQ}}$ in the above
formulas defined for $\eta_{PQ}=-1$ are
\begin{eqnarray}
\mbox{if} \ \theta_{PQ} < 0&:& \\ \nonumber
\Psi(\theta^j_{PQ})&=&
-\gamma_E+\frac{1}{2}\frac{d}{d{\theta}^j_{PQ}}\,\ln{\Gamma(-\theta^j_{PQ})}-
\frac{1}{2}\frac{d}{d{\theta}^j_{PQ}}\,\ln{\Gamma(1+\theta^j_{PQ})}+\ln(t^j+\bar t^j)\\ \nonumber
\mbox{if} \ \theta_{PQ} > 0&:& \\ \nonumber
\Psi(\theta^j_{PQ})&=&
-\gamma_E+\frac{1}{2}\frac{d}{d{\theta}^j_{PQ}}\,\ln{\Gamma(1-\theta^j_{PQ})}-
\frac{1}{2}\frac{d}{d{\theta}^j_{PQ}}\,\ln{\Gamma(\theta^j_{PQ})}+\ln(t^j+\bar t^j),
\label{eqn:Psi1}
\end{eqnarray}
and for $\eta_{PQ}=+1$ are
\begin{eqnarray}
\mbox{if} \ \theta_{PQ} < 0&:& \\ \nonumber
\Psi(\theta^j_{PQ})&=&
\gamma_E+\frac{1}{2}\frac{d}{d{\theta}^j_{PQ}}\,\ln{\Gamma(1+\theta^j_{PQ})}-
\frac{1}{2}\frac{d}{d{\theta}^j_{PQ}}\,\ln{\Gamma(-\theta^j_{PQ})}-\ln(t^j+\bar t^j)\\ \nonumber
\mbox{if} \ \theta_{PQ} > 0&:& \\ \nonumber
\Psi(\theta^j_{PQ})&=&
\gamma_E+\frac{1}{2}\frac{d}{d{\theta}^j_{PQ}}\,\ln{\Gamma(\theta^j_{PQ})}-
\frac{1}{2}\frac{d}{d{\theta}^j_{PQ}}\,\ln{\Gamma(1-\theta^j_{PQ})}-\ln(t^j+\bar t^j).
\label{eqn:Psi2}
\end{eqnarray}
The function $\Psi'(\theta_{PQ})$ is just the derivative
\begin{eqnarray}
\Psi'(\theta^j_{PQ})&=&
\frac{d\Psi(\theta^j_{PQ})}{d \theta^j_{PQ}},
\label{eqn:Psip}
\end{eqnarray}
and ${\theta}^{j,k}_{PQ}$ and ${\theta}^{j,k\bar{l}}_{PQ}$ are
defined~\cite{Kane:2004hm} as

\begin{equation}
{\xi}^{j,k}_{PQ} \equiv (u^k+\bar u^k)\,\frac{\partial
\theta^j_{PQ}}{\partial u^k}= \left\{\begin{array}{cc}
 \left[-\frac{1}{4\pi}
 \sin(2\pi\theta^j)
 \right]^P_Q & \mbox{ when }j=k  \vspace*{0.6cm} \\
 \left[\frac{1}{4\pi}
\sin(2\pi\theta^j)
 \right]^P_Q & \mbox{ when }j\neq k,
\end{array}\right.\label{idb:eq:dthdu}
\end{equation}

\begin{equation}
{\xi}^{j,k\bar{l}}_{PQ} \equiv (u^k+\bar
u^k)(u^l+\bar u^l)\,\frac{\partial^2 \theta^j_{PQ}}{\partial
u^k\partial\bar u^l}= \left\{\begin{array}{cc} \frac{1}{16\pi}
  \left[ \sin(4\pi\theta^j)+4\sin(2\pi\theta^j)
 \right]^P_Q &
   \mbox{when }j=k=l  \vspace*{0.6cm} \\
 \frac{1}{16\pi}  \left[
 \sin(4\pi\theta^j)-4\sin(2\pi\theta^j)
 \right]^P_Q &
   \mbox{when }j\neq k=l  \vspace*{0.6cm} \\
 -\frac{1}{16\pi}\left[
 \sin(4\pi\theta^j)
 \right]^P_Q &
   \mbox{ when }j=k\neq l\mbox{ or } j=l\neq k \vspace*{0.4cm} \\
 \frac{1}{16\pi}\left[
\sin(4\pi\theta^j)
 \right]^P_Q &
   \mbox{when }j\neq k\neq l\neq j.
\end{array}\right.\label{idb:eq:dth2du}
\end{equation}

Note that the only explicit dependence of the soft terms on the
$u$ and $s$ moduli is in the gaugino mass parameters.  The
trilinears and scalar mass-squared values depend explicitly only
on the angles.  However, there is an implicit dependence on the
complex structure moduli via the angles made by each D-brane with respect to the
orientifold planes.

In contrast to heterotic string models, the gaugino and scalar
masses are typically not universal in intersecting D-brane
constructions, although in the present case, there is some partial
universality of the scalar masses due to the Pati-Salam
unification at the string scale. In particular, the trilinear $A$
couplings are found to be equal to a universal parameter, $A_0$,
and the left-handed and right-handed squarks and sleptons
respectively are degenerate. The Higgs states arise from the
non-chiral sector due to the fact that stacks $b$, $c1$, and $c2$
are parallel on the third torus.  The appropriate K\a"ahler metric
for these states is given by Eq.~(\ref{nonchiralK}).  Thus, the
Higgs scalar mass-squared values are found to be
\begin{equation}
m^2_H = m^2_{3/2}\left(1-\frac{3}{2}\left|\Theta_3\right|^2\right).
\label{eq:idb:higgsmass_u}
\end{equation}

The complex structure moduli $u^i$ and the four-dimensional
dilaton $\phi_4$ are fixed by the supersymmetry conditions and
gauge coupling unification, respectively. The K\a"ahler modulus on
the first torus $t^1$ will be chosen to be consistent with the Yukawa mass
matrices. Thus, the free parameters which remain are
$\Theta_{1}$, $\Theta_{2}$, $\mbox{sgn}(\Theta_3)$, $t^2$, $t^3$, the phases
$\gamma_i$, and the gravitino mass $m_{3/2}$. In order to
eliminate potential problems with electric dipole moments (EDM's) for the 
neutron and electron, we set $\gamma_i=0$. In
addition, we set the K\a"ahler moduli on the second and third tori
equal to one another, $\mbox{Re}(t^2)=\mbox{Re}(t^3)=0.5$. Note that the soft terms
only have a weak logarithmic dependence on the K\a"ahler moduli. We constrain the parameter space such that
neither the Higgs nor the squark and slepton scalar masses are
tachyonic at the high scale, as well as imposing the unitary condition
$\Theta_1^2+\Theta_2^2+\Theta_3^3=1$. In particular, we require
$\Theta_3^2 \leq 2/3$, or equivalently $\Theta_1^2+\Theta_2^2 \geq
1/3$. We thus now have three free parameters, $\Theta_{1}$, $\Theta_{2}$, and $m_{3/2}$.

Our goal in this work is to construct the expected final states at LHC and discuss how the model parameters can be determined at LHC for an intersecting $D$6-brane model. First, we generate sets of soft-supersymmetry breaking terms to reveal those regions of the parameter space that satisfy all the presently known experimental constraints and can generate the WMAP observed dark matter density and the relic density in the SSC scenario. Then we categorize all the regions of the experimentally allowed parameter space into different patterns of the superpartner mass spectra, where these patterns are organized by the masses of the four lightest sparticles. Using this data, we construct the intersecting $D$6-brane model  final states at LHC and compare to the final states of mSUGRA. Next we show that the correct dark matter density can be obtained within this model in both the stau-neutralino and chargino-neutralino coannihilation regions. Finally, we discuss the challenges of constructing experimental observables to determine the $D$6-brane model parameters.

\section{Parameter Space and Supersymmetry Spectra}

We generate sets of the seven soft-supersymmetry breaking mass parameters at the unification scale using the equations given in Eqs.~(\ref{eq:idb:gaugino}),~(\ref{eq:idb:bino}),~(\ref{eq:idb:trilinear_u}),~(\ref{eq:idb:scalarmass_u}), and~(\ref{eq:idb:higgsmass_u}) for $\textit{u}$-moduli dominated SUSY breaking. The seven soft-supersymmetry breaking mass parameters are the gaugino masses $M3$, $M2$, and $M1$, the Higgs scalar mass-squared $m_{H}^{2}$, the left scalar mass $m_{L}$, the right scalar mass $m_{R}$, and the universal trilinear coupling $A_{0}$. We leave tan$\beta$ as a free parameter, which gives a total of four free parameters, $\Theta_{1},~ \Theta_{2},~ m_{3/2}$, and tan$\beta$, so we are led to a four-parameter model. The seven soft-supersymmetry breaking mass parameters at the unification scale are functions of the three goldstino angles $\Theta_{1},~ \Theta_{2},~ m_{3/2}$ which parameterize the F-terms.

The parameters are input into {\tt MicrOMEGAs 2.0.7}~\cite{Belanger:2006is} using {\tt SuSpect 2.34}~\cite{Djouadi:2002ze} as a front end running the soft terms down to the electroweak scale via the Renormalization Group Equations (RGEs) to calculate the supersymmetry particle spectra and then to calculate the corresponding relic neutralino density.  We take the top quark mass to be $m_t = 172.6$ GeV~\cite{:2008vn}, while $\mu$ is determined by the requirement of radiative electroweak symmetry breaking (REWSB). However, we do take $\mu > 0$ as suggested by the results of $g_{\mu}-2$ for the muon. The resulting superpartner spectra are then filtered according to the following criteria: 

\begin{enumerate}

\item The 5-year WMAP data combined with measurements of Type Ia supernovae and baryon acoustic oscillations in the galaxy distribution for the cold dark matter density~\cite{Hinshaw:2008kr},  0.1109 $\leq \Omega_{\chi^o} h^{2} \leq$ 0.1177, where a neutralino LSP is the dominant component of the relic density. In addition, we look at the SSC model~\cite{Antoniadis:1988aa}, in which a dilution factor of $\cal{O}$(10) is allowed~\cite{Lahanas:2006hf}, where $\Omega_{\chi^o} h^{2} \lesssim$ 1.1. For a discussion of the SSC model within the context of mSUGRA, see~\cite{Dutta:2008ge}. We also investigate another case where a neutralino LSP makes up a subdominant component, allowing for the possibility that dark matter could be composed of matter such as axions, cryptons, or other particles. We employ this possibility by removing the lower bound.

\item The experimental limits on the Flavor Changing Neutral Current (FCNC) process, $b \rightarrow s\gamma$. The results from the Heavy Flavor Averaging Group (HFAG)~\cite{Barberio:2007cr}, in addition to the BABAR, Belle, and CLEO results, are: $Br(b \rightarrow s\gamma) = (355 \pm 24^{+9}_{-10} \pm 3) \times 10^{-6}$. There is also a more recent estimate~\cite{Misiak:2006zs} of $Br(b \rightarrow s\gamma) = (3.15 \pm 0.23) \times 10^{-4}$. For our analysis, we use the limits $2.86 \times 10^{-4} \leq Br(b \rightarrow s\gamma) \leq 4.18 \times 10^{-4}$, where experimental and
theoretical errors are added in quadrature.

\item The anomalous magnetic moment of the muon, $g_{\mu} - 2$. For this analysis we use the 2$\sigma$ level boundaries, $11 \times 10^{-10} < a_{\mu} < 44 \times 10^{-10}$~\cite{Bennett:2004pv}.

\item The process $B_{s}^{0} \rightarrow \mu^+ \mu^-$ where the decay has a $\mbox{tan}^6\beta$ dependence. We take the upper bound to be $Br(B_{s}^{0} \rightarrow \mu^{+}\mu^{-}) < 5.8 \times 10^{-8}$~\cite{:2007kv}.

\item The LEP limit on the lightest CP-even Higgs boson mass, $m_{h} \geq 114$ GeV~\cite{Barate:2003sz}.

\end{enumerate}

The gravitino mass $m_{3/2}$ linearly scales the seven mass parameters at the unification scale. We scan these seven mass parameters for the $\textit{u}$-moduli dominated SUSY breaking scenario for various values of $m_{3/2}$ and tan$\beta$ to determine a suitable range for $m_{3/2}$, where we want to establish an upper limit such that $m_{3/2}$ becomes too massive at which few sparticles could be produced at LHC for an integrated luminosity of 10 $fb^{-1}$, and at the lower limit the Higgs mass becomes too light and violates the LEP constraint. To satisfy these conditions, we position the upper limit to be $m_{3/2}$ $\approx$ 700 GeV and compute the lower limit to be in the range $m_{3/2}$ = 400 $\sim$ 500 GeV. Consequently, to efficiently execute the substantial quantity of requisite computations, we limit our calculations of the experimental constraints, supersymmetry spectra, and relic density to $m_{3/2}$ = 500 GeV and $m_{3/2}$ = 700 GeV. For each $m_{3/2}$, the calculations were completed for tan$\beta$ = 10, 25, and 46. Regions of the parameter space satisfying all the experimental constraints exist for five of the six cases; only $m_{3/2}$ = 700 GeV, tan$\beta$ = 10 produced no spectra that fulfilled the constraints. Additional low values of tan$\beta$ were run for $m_{3/2}$ = 700 GeV, though tan$\beta$ = 25 is the approximate minimum tan$\beta$ that violates none of the constraints. Thus, we study five cases for the $\textit{u}$-moduli dominated SUSY breaking scenario in this work: $m_{3/2}$ = 500 GeV and tan$\beta$ = 10, $m_{3/2}$ = 500 GeV and tan$\beta$ = 25, $m_{3/2}$ = 500 GeV and tan$\beta$ = 46, $m_{3/2}$ = 700 GeV and tan$\beta$ = 25, $m_{3/2}$ = 700 GeV and tan$\beta$ = 46. These five cases will produce a broad spectrum of mass parameters at the unification scale such that a representative allowed parameter space can be determined.

\begin{figure}[htp]
	\centering
		\includegraphics[width=0.38\textwidth]{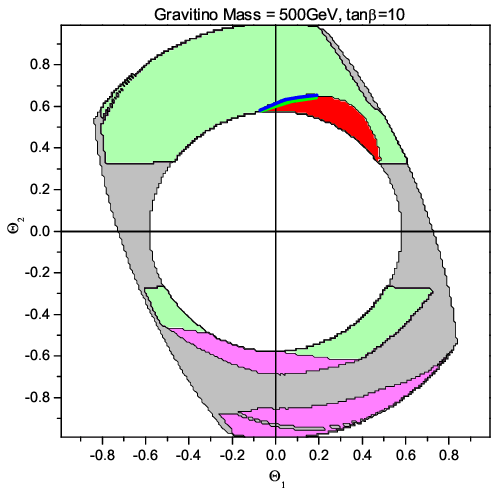}
		\includegraphics[width=0.38\textwidth]{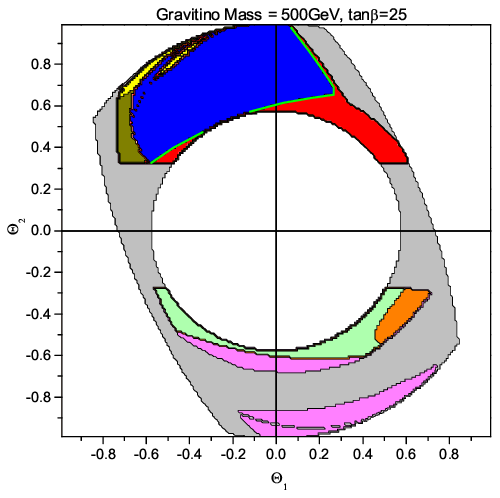}
		\includegraphics[width=0.38\textwidth]{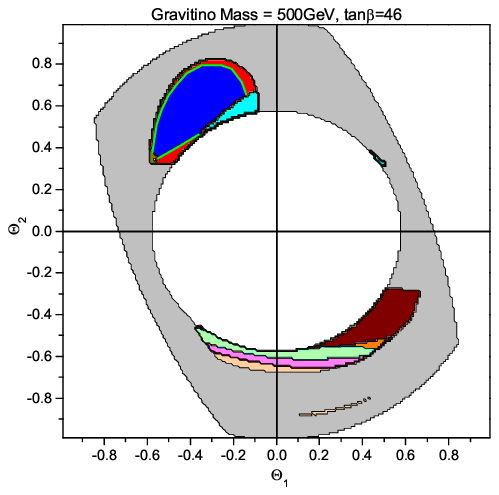}
		\includegraphics[width=0.38\textwidth]{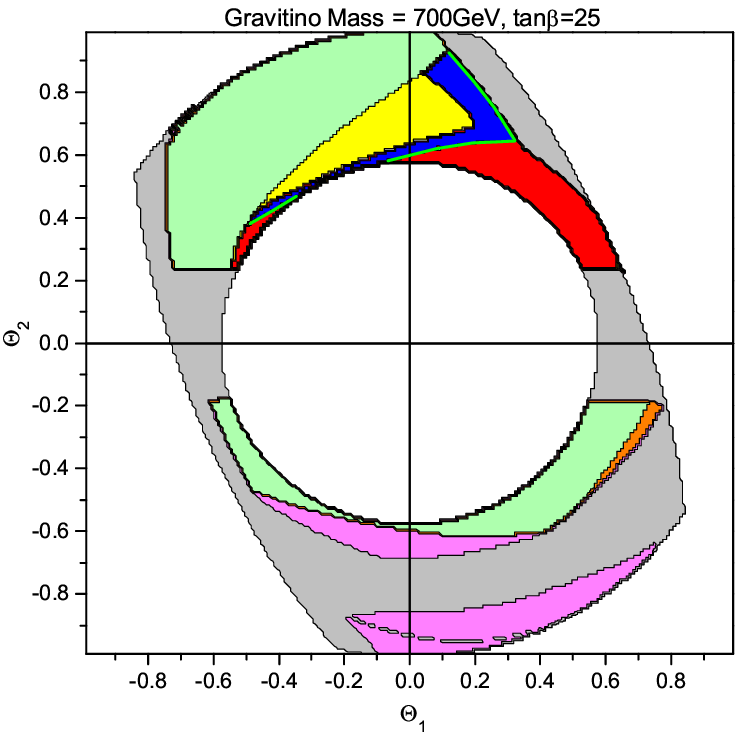}
		\includegraphics[width=0.38\textwidth]{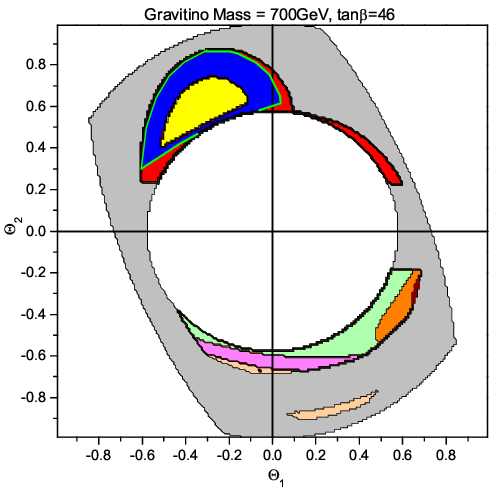}
		\includegraphics[width=0.38\textwidth]{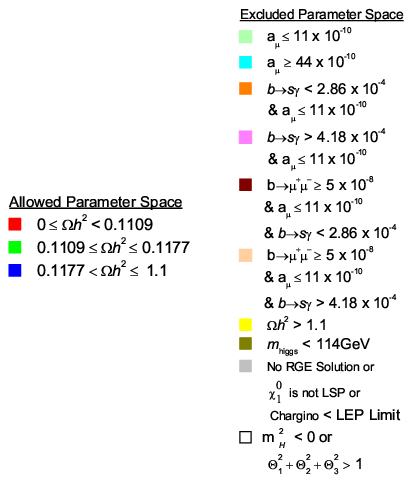}
		\caption{Allowed parameter space for $\textit{u}$-moduli dominated SUSY breaking scenario for an intersecting $D$6-brane model. The five individual charts represent different gravitino masses and tan$\beta$. The chart legend describes the reasons for inclusion and exclusion of the shaded regions. Each separate region is outlined in black. Note the small regions excluded by the Higgs mass $m_h$ $<$ 114 GeV and $\Omega_{\chi^o} h^{2} >$ 1.1 satisfy all other constraints. The unshaded circular region centered at the origin is prohibited for driving $m_{H}^{2}$ to negative values, and the remaining unshaded regions are rejected since $\Theta_{1}^{2} + \Theta_{2}^{2} + \Theta_{3}^{2} \neq 1$.}
	\label{fig:D6_ParamSpace}
\end{figure}

We delineate the parameter space for the $D$6-brane model in terms of the goldstino angles $\Theta_{1}$ and $\Theta_{2}$. For clarity, we segregate the parameter space into distinctive scenarios of $m_{3/2}$ and tan$\beta$, each scenario delineated by $\Theta_{1}$ and $\Theta_{2}$. One set of these four free parameters determines a unique point in the parameter space described by the seven soft-supersymmetry breaking mass parameters. The experimentally allowed parameter space for each of the five scenarios of $m_{3/2}$ and tan$\beta$ is exhibited in Fig.~\ref{fig:D6_ParamSpace}. Note in Fig.~\ref{fig:D6_ParamSpace} that very constrained regions in the allowed parameter space exist that can generate the WMAP observed dark matter density, and furthermore, larger regions exist that can generate the diluted relic density in the SSC scenario. We see one consequence of raising $m_{3/2}$, which in effect increases the mass parameters, most consistently the gaugino mass $M3$, is to drive the relic density of some regions with already high levels of $\Omega_{\chi}$ to levels where $\Omega_{\chi} \geq 1.1$. The increase in the mass parameters expands the mass difference between the lightest SUSY particle (LSP) $\widetilde{\chi}_{1}^{0}$ and the next to lightest SUSY particle (NLSP), thereby diminishing the prospects for coannihilation between the LSP and NLSP, and as a result, elevating the relic density. Those regions in Fig.~\ref{fig:D6_ParamSpace} that can generate the WMAP observed dark matter density and relic density in the context of SSC are vital in this work to uncovering the expected final states at LHC.

We find that different regions of the parameter space that are allowed by the experimental constraints possess different patterns of mass hierarchies of the four lightest supersymmetric partners. Identification of the landscape of such mass patterns is of interest in classifying the possible experimental signals that may be expected at LHC~\cite{Feldman:2007zn}. Through a comprehensive scan of all regions of the allowed parameter space, we uncover five such patterns of mass hierarchies present.  The five patterns present in the supersymmetry parameter space are shown in Table~\ref{tab:Masspatt} for the $\textit{u}$-moduli dominated SUSY breaking scenario.

\begin{table}[ht]
	\centering
	\caption{Patterns of the four lightest sparticles for spectra allowed by all constraints for the intersecting $D6$-brane model (IBM).}
		\begin{tabular}{|c|c|c|c|} \hline
		$\textnormal{Model}$ & $\textnormal{Pattern No.}$ &  $\textnormal{Pattern Type}$ & $\textnormal{Mass Pattern}$ \\ \hline\hline
		$\textnormal{IBM}$  &  $\textnormal{\it{ID6BraneP1}}$  &  $\textnormal{Chargino}$  &  $~\widetilde{\chi}_{1}^{0} ~<~ \widetilde{\chi}_{1}^{\pm} ~<~ \widetilde{\chi}_{2}^{0} ~<~ \widetilde{\tau}~$\\ \hline
		$\textnormal{IBM}$  &  $\textnormal{\it{ID6BraneP2}}$  &  $\textnormal{Chargino}$  &  $~\widetilde{\chi}_{1}^{0} ~<~ \widetilde{\chi}_{1}^{\pm} ~<~ \widetilde{\tau} ~<~ \widetilde{\chi}_{2}^{0}~$\\	\hline
		$\textnormal{IBM}$  &  $\textnormal{\it{ID6BraneP3}}$  &  $\textnormal{Chargino}$  &  $~\widetilde{\chi}_{1}^{0} ~<~ \widetilde{\chi}_{1}^{\pm} ~<~ \widetilde{\tau} ~<~ \widetilde{e}_{R}~$\\	\hline
		$\textnormal{IBM}$  &  $\textnormal{\it{ID6BraneP4}}$  &  $\textnormal{Stau}$  &  $~\widetilde{\chi}_{1}^{0} ~<~ \widetilde{\tau} ~<~ \widetilde{\chi}_{1}^{\pm} ~<~ \widetilde{\chi}_{2}^{0}~$\\	\hline
		$\textnormal{IBM}$  &  $\textnormal{\it{ID6BraneP5}}$  &  $\textnormal{Stau}$  &  $~\widetilde{\chi}_{1}^{0} ~<~ \widetilde{\tau} ~<~ \widetilde{e}_{R} ~<~ \widetilde{\chi}_{1}^{\pm}~$\\ \hline
		\end{tabular}
		\label{tab:Masspatt}
\end{table}

We now discuss each of these five patterns in detail. The $\widetilde{\chi}_{1}^{\pm}$ is the NLSP for the first three patterns in Table~\ref{tab:Masspatt}. A small region of the allowed parameter space with the $\textit{ID6BraneP1}$ pattern has the $\widetilde{\chi}_{1}^{\pm}$ and $\widetilde{\chi}_{2}^{0}$ mass nearly degenerate with the $\widetilde{\chi}_{1}^{0}$, with a mass difference $\lesssim$ 20 GeV, allowing for the observed dark matter density by WMAP to be generated in the chargino-neutralino coannihilation region. In addition, a large region of the allowed parameter space with the $\textit{ID6BraneP1}$ pattern has a very large mass difference between the $\widetilde{\chi}_{1}^{\pm}$ and $\widetilde{\chi}_{1}^{0}$, up to $\sim$150 GeV, generating a dark matter density up to $\Omega_{\chi}\sim$1.1, possessing characteristics of the SSC scenario. In those regions of the parameter space in the SSC scenario with pattern $\textit{ID6BraneP1}$, the $\widetilde{\chi}_{1}^{\pm}$ and $\widetilde{\chi}_{2}^{0}$ are virtually degenerate as well. We shall discuss the case of neutralino coannihilation in more detail later. 

All the regions of the allowed parameter space with patterns $\textit{ID6BraneP2}$ and $\textit{ID6BraneP3}$ have a virtually degenerate mass between the $\widetilde{\chi}_{1}^{\pm}$ and $\widetilde{\chi}_{1}^{0}$, with a mass difference of less than 1 GeV. The virtually degenerate mass between the $\widetilde{\chi}_{1}^{\pm}$ and $\widetilde{\chi}_{1}^{0}$ in the regions of the allowed parameter space with patterns $\textit{ID6BraneP2}$ and $\textit{ID6BraneP3}$ allow for only a very small dark matter density to be generated, $\Omega_{\chi} \lesssim 0.01$, well below the WMAP observed relic density. Thus, in these regions with patterns $\textit{ID6BraneP2}$ and $\textit{ID6BraneP3}$, the WMAP observed dark matter density must be predominantly composed of something other than the LSP since the lightest neutralino can only generate a small fraction of the dark matter for these regions of the parameter space. 

We identify the fourth and fifth patterns in Table~\ref{tab:Masspatt} as stau patterns since the $\widetilde{\tau}_{1}$ is the NLSP. As we shall soon discuss, regions of the allowed parameter space with the $\textit{ID6BraneP4}$ and $\textit{ID6BraneP5}$ patterns in the intersecting $D$6-brane model will produce physics similar to the stau-neutralino coannihilation region in mSUGRA. There are small regions of the allowed parameter space with both the $\textit{ID6BraneP4}$ and $\textit{ID6BraneP5}$ patterns with a mass difference between the $\widetilde{\tau}_{1}$ and $\widetilde{\chi}_{1}^{0}$ less than $\sim$20 GeV, generating the WMAP observed dark matter density in the stau-neutralino coannihilation region. Furthermore, there are large regions of the allowed parameter space with patterns $\textit{ID6BraneP4}$ and $\textit{ID6BraneP5}$ that have a mass difference between the $\widetilde{\tau}_{1}$ and $\widetilde{\chi}_{1}^{0}$ of up to $\sim$160 GeV, generating a dark matter density up to $\Omega_{\chi}\sim$1.1, within the SSC scenario. 

\begin{figure}[htp]
  \centering
		\includegraphics[width=0.41\textwidth]{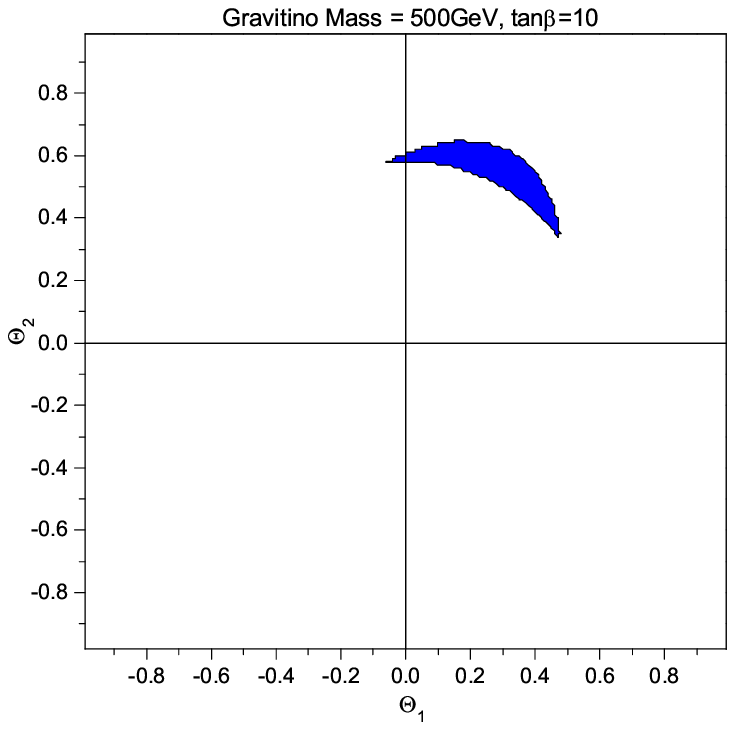}
		\includegraphics[width=0.41\textwidth]{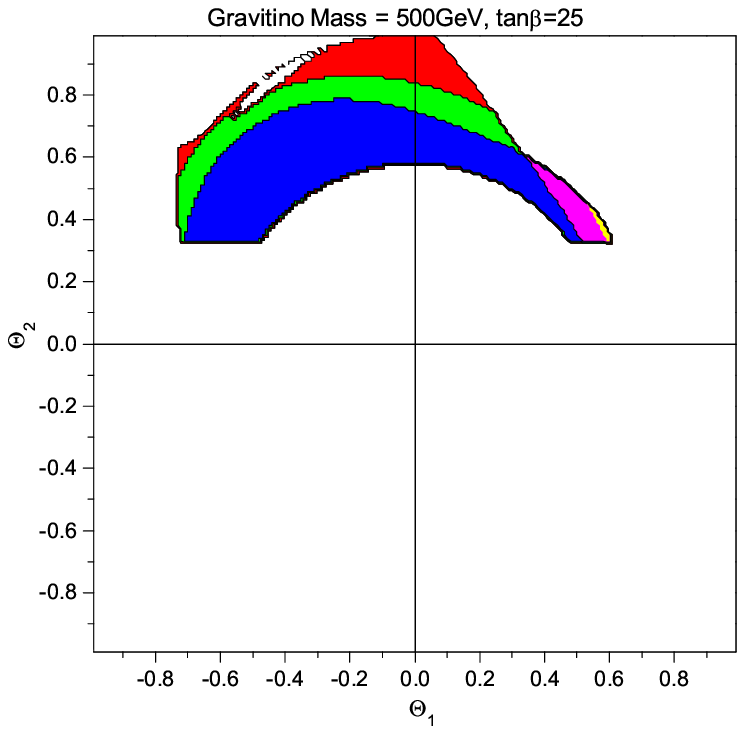}
		\includegraphics[width=0.41\textwidth]{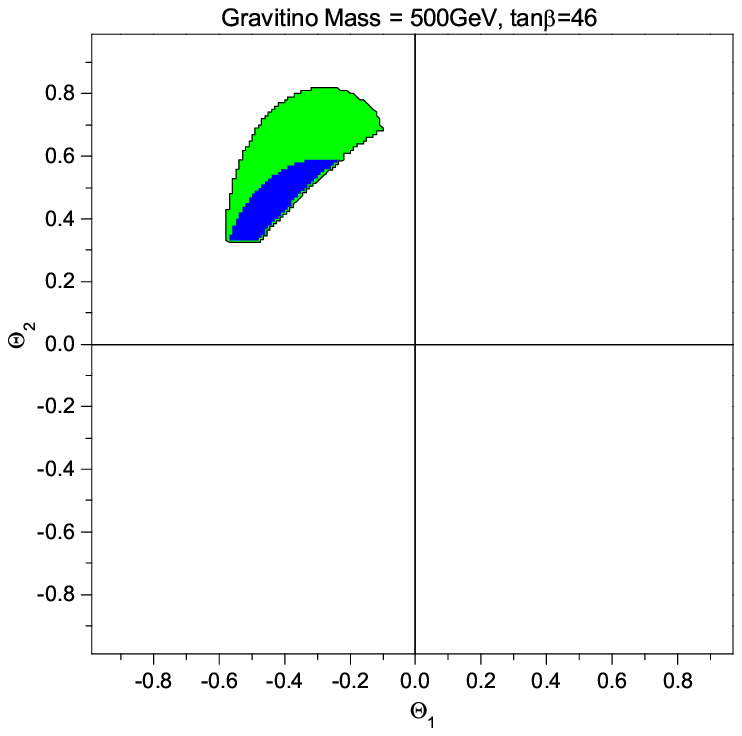}
		\includegraphics[width=0.41\textwidth]{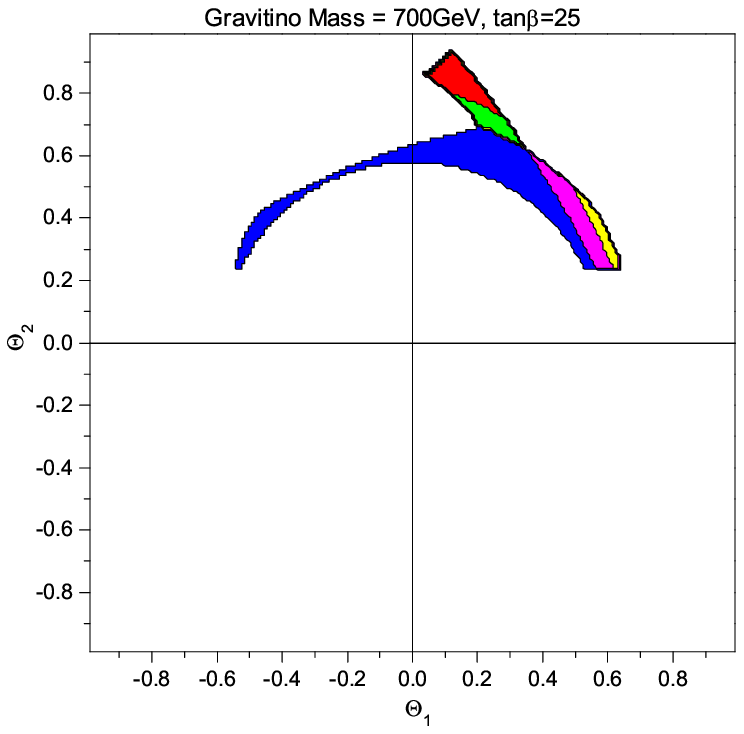}
		\includegraphics[width=0.41\textwidth]{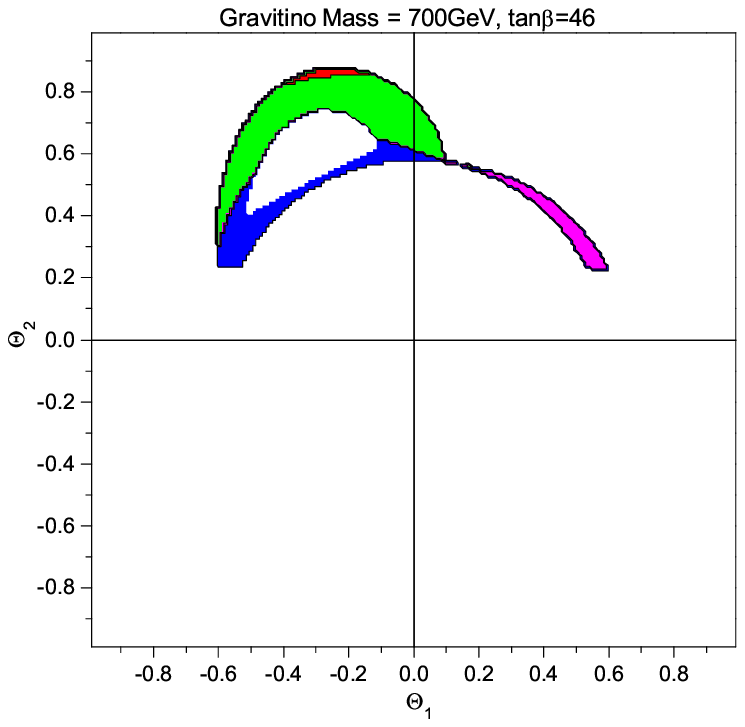}
		\includegraphics[width=0.41\textwidth]{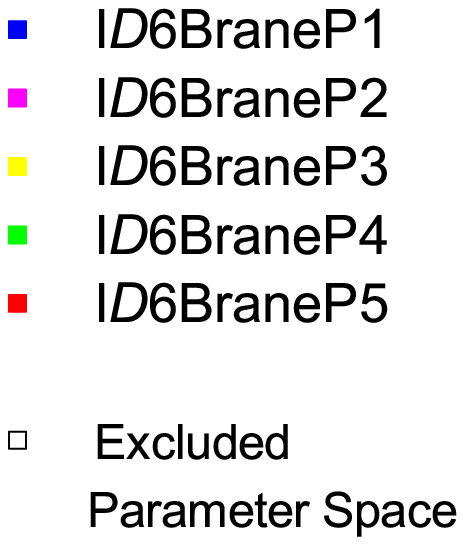}
		\caption{Patterns of the mass spectra allowed by all the experimental constraints for the $\textit{u}$-moduli dominated SUSY breaking scenario. The five individual charts represent different gravitino masses and tan$\beta$. The allowed parameter space here correlates directly with the allowed parameter space in Fig~\ref{fig:D6_ParamSpace}. The shaded regions within each chart identify the five different patterns, and each separate region is outlined in black.}
	\label{fig:D6_MassSpace}
\end{figure}

When discussing the final states which may be produced at the LHC in the next section, we shall focus on the $\textit{ID6BraneP1}$, $\textit{ID6BraneP4}$, and $\textit{ID6BraneP5}$ patterns since only these three patterns can generate the WMAP observed relic density, within the chargino-neutralino and stau-neutralino coannihilation regions, in addition to the diluted dark matter density in the context of SSC. We show in Fig.~\ref{fig:D6_MassSpace} all the regions of the allowed parameter space partitioned into the five patterns of the mass spectra we have discussed. In order to correlate the pattern space in Fig.~\ref{fig:D6_MassSpace} with the allowed parameter space in Fig.~\ref{fig:D6_ParamSpace}, the plots of the different patterns of the mass spectra in Fig.~\ref{fig:D6_MassSpace} are also delineated in terms of $\Theta_{1}$ and $\Theta_{2}$, segregated into the five $m_{3/2}$ and tan$\beta$ scenarios for clarity. The parameter space shown in Fig.~\ref{fig:D6_ParamSpace} and the correlated landscape of mass patterns shown in Fig.~\ref{fig:D6_MassSpace} will serve as the basis for selection of typical points with which to derive the final states at LHC in the next section.

\section{The Final States at LHC}

The ultimate goal is to derive the model parameters from the experimental data.  This is accomplished by constructing experimental kinematic observables that extract the expected final states while suppressing the Standard Model background. Measurements of the kinematic variables are then used to compute the model parameters. The final states in the present model will vary dependent upon the superpartner mass spectra patterns we have identified since each pattern may possess distinctive dominant decay chains and final states. Only regions of the parameter space with patterns $\textit{ID6BraneP1}$, $\textit{ID6BraneP4}$, and $\textit{ID6BraneP5}$ will generate the WMAP observed relic density and the diluted dark matter density in the SSC scenario, so consequently, we shall only analyze the final states at LHC for points within the regions of the allowed parameter space with these three patterns. We select typical points from each of these three regions of the parameter space and examine the decay modes and final states. We shall choose a sample point and compute the final states, then vary the gaugino mass parameters $M3$, $M2$, and $M1$, the Higgs scalar mass-squared parameter $m_{H}^{2}$, the left scalar mass parameter $m_{L}$, the right scalar mass parameter $m_{R}$, and the universal trilinear coupling parameter $A_{0}$ to understand the effect of the variance on the states, while leaving tan$\beta$ constant. The cross-sections and branching ratios are calculated with PYTHIA 6.411~\cite{Sjostrand:2006za}, using {\tt SuSpect 2.34} to compute the sparticle masses.

We first analyze points within regions of the parameter space with pattern $\textit{ID6BraneP1}$ that generate the WMAP observed dark matter density. These points reside in the chargino-neutralino coannihilation region due to the small mass difference between the $\widetilde{\chi}_{1}^{\pm}$ and $\widetilde{\chi}_{2}^{0}$ with the $\widetilde{\chi}_{1}^{0}$. The processes with the largest production cross-sections are $q+\overline{q}\rightarrow\widetilde{\chi}_{2}^{0}+\widetilde{\chi}_{1}^{\pm}$ and $q+\overline{q}\rightarrow\widetilde{q}+\widetilde{q}$. The NLSP is the $\widetilde{\chi}_{1}^{\pm}$, which is virtually degenerate with the $\widetilde{\chi}_{2}^{0}$, so for this reason, the $\widetilde{\chi}_{1}^{\pm}$ and $\widetilde{\chi}_{2}^{0}$ have large production cross-sections. Recall $\left\langle \sigma_{\textit{ann}}\textit{v}\right\rangle \propto \frac{1}{m^{2}}$, thus the more massive the particle, the smaller the differential cross-section. Since the $\widetilde{\tau}_{1}^{\pm}$ is more massive than the either the $\widetilde{\chi}_{1}^{\pm}$ or $\widetilde{\chi}_{2}^{0}$, the $\widetilde{\chi}_{1}^{\pm}$ and $\widetilde{\chi}_{2}^{0}$ will decay directly to opposite sign $\sim$20GeV lepton pairs and hadronic jets. Two typical points are shown in Table~\ref{tab:P1_WMAP_1} and Table~\ref{tab:P1_WMAP_2}. We see the most favored decay for $\widetilde{\chi}_{2}^{0}$ is $\widetilde{\chi}_{2}^{0} \rightarrow \nu\overline{\nu}\widetilde{\chi}_{1}^{0}$, producing high energy $\sim$20GeV neutrinos. This is certainly not a dominant decay mode since production of low energy leptons $(e,\mu,\tau)$ have a roughly equal branching ratio to the production of these high energy neutrinos. The $\widetilde{\chi}_{1}^{0}$ and neutrinos will exit the detector undetected, producing only missing energy $\mbox{E\!\!\!\!/}_{T}$. For the decay of $\widetilde{\chi}_{1}^{\pm}$, we are chiefly looking at the production of jets through $\widetilde{\chi}_{1}^{\pm} \rightarrow q\overline{q}\widetilde{\chi}_{1}^{0}$, with a smaller branching ratio for the decay to low energy leptons. The other primary sources of jets are from 
$\widetilde{q}_{R} \rightarrow q\widetilde{\chi}_{1}^{0}$, $\widetilde{q}_{L} \rightarrow q\widetilde{\chi}_{2}^{0}$, and $\widetilde{q}_{L} \rightarrow q\widetilde{\chi}_{1}^{\pm}$. There is no change in these decay modes when we vary the mass parameters. Therefore, we have three principal signals to expect at LHC for these points that produce the WMAP observed dark matter density, where $l = (e,\mu)$:

\begin{itemize}
 
  \item $jets + \mbox{E\!\!\!\!/}_{T}$
  \item $2\tau + jets + \mbox{E\!\!\!\!/}_{T}$
  \item $2l + jets + \mbox{E\!\!\!\!/}_{T}$

\end{itemize}

Now we examine points within regions of the parameter space with pattern $\textit{ID6BraneP1}$ that generate the diluted dark matter density in the context of SSC. The mass difference between the $\widetilde{\chi}_{1}^{\pm}$ and $\widetilde{\chi}_{2}^{0}$ with the $\widetilde{\chi}_{1}^{0}$ is much greater, so these points do not necessarily lie within the chargino-neutralino coannihilation region of the parameter space. The three reference points we select are shown in Tables~\ref{tab:P1_SSC_1},~\ref{tab:P1_SSC_2}, and~\ref{tab:P1_SSC_3}. The processes with the largest production cross-sections for the point shown in Table~\ref{tab:P1_SSC_1} are $q+\overline{q}\rightarrow\widetilde{\chi}_{1}^{\pm}+\widetilde{\chi}_{1}^{\pm}$ and $q+\overline{q}\rightarrow\widetilde{q}+\widetilde{q}$. This point has the smallest mass parameters of these three points, however, as we increase the mass parameters to those in Tables~\ref{tab:P1_SSC_2} and~\ref{tab:P1_SSC_3}, the production cross-sections for the aforementioned processes remain large, though the largest cross-section becomes $q+\overline{q}\rightarrow\widetilde{\chi}_{1}^{0}+\widetilde{\chi}_{1}^{\pm}$. We have the same dominant decay modes as the WMAP points, but as the mass parameters increase, the branching ratios for the $\widetilde{\chi}_{1}^{\pm}$ decays change only slightly, while the branching ratio for the pair of high energy neutrinos increases to as high as 50\%. Increasing the mass parameters also decreases the number of $q\overline{q}$ jets produced from $\widetilde{\chi}_{2}^{0}$. Additionally, the branching ratios for the production of two tau decrease from those of the WMAP regions, and the larger mass difference between the $\widetilde{\chi}_{1}^{\pm}$ and $\widetilde{\chi}_{2}^{0}$ with the $\widetilde{\chi}_{1}^{0}$ will produce $\gtrsim$ 20 GeV lepton pairs and neutrinos. The primary source of jets are the same as the WMAP regions. Hence, we expect essentially the same signals as those of the WMAP regions, but the signals in the SSC regions should be easily distinguished from the WMAP regions by observing a larger mass difference between the $\widetilde{\chi}_{1}^{\pm}$ and $\widetilde{\chi}_{2}^{0}$ with the $\widetilde{\chi}_{1}^{0}$.

\begin{table}[t]
\caption{Low energy supersymmetric particles and masses (in GeV) for $\textit{ID6BraneP1}$ point, $\Theta_{1}$ = -0.08, $\Theta_{2}$ = 0.58, $M3$ = 602, $M2$ = 251, $M1$ = 430, $m_{H}$ = 59, $m_{L}$ = 273, $m_{R}$ = 312, $A_{0}$ = -37, tan$\beta$ = 25, $m_{3/2}$ = 500. The relic density for this point is $\Omega_{\chi}$ = 0.1127. Here, $l = (e,\mu)$.}
 \label{tab:P1_WMAP_1}
\begin{center}
\begin{tabular}{c c c c c c c c c c}
\hline \hline
$\widetilde{g}$ &
$\begin{array}{c} \widetilde{u}_{L} \\ \widetilde{u}_{R} \end{array}$ &
$\begin{array}{c} \widetilde{t}_{2} \\ \widetilde{t}_{1} \end{array}$ &
$\begin{array}{c} \widetilde{b}_{2} \\ \widetilde{b}_{1} \end{array}$ &
$\begin{array}{c} \widetilde{e}_{L} \\ \widetilde{e}_{R} \end{array}$ &
$\begin{array}{c} \widetilde{\tau}_{2} \\ \widetilde{\tau}_{1} \end{array}$ &
$\begin{array}{c} \widetilde{\chi}_{2}^{0} \\ \widetilde{\chi}_{1}^{0} \end{array}$ &
$\begin{array}{c} \widetilde{\chi}_{2}^{\pm} \\ \widetilde{\chi}_{1}^{\pm} \end{array}$ &
$\begin{array}{c} Br(\widetilde{\chi}_{2}^{0} \rightarrow q\overline{q}\widetilde{\chi}_{1}^{0})(\%) \\ Br(\widetilde{\chi}_{2}^{0} \rightarrow l^{+}l^{-}\widetilde{\chi}_{1}^{0})(\%) \\ Br(\widetilde{\chi}_{2}^{0} \rightarrow \tau^{+}\tau^{-}\widetilde{\chi}_{1}^{0})(\%) \\ Br(\widetilde{\chi}_{2}^{0} \rightarrow \nu\overline{\nu}\widetilde{\chi}_{1}^{0})(\%) \end{array}$ &
$\begin{array}{c} Br(\widetilde{\chi}_{1}^{\pm} \rightarrow q\overline{q}\widetilde{\chi}_{1}^{0})(\%) \\ Br(\widetilde{\chi}_{1}^{\pm} \rightarrow l^{\pm}\nu\widetilde{\chi}_{1}^{0})(\%) \\ Br(\widetilde{\chi}_{1}^{\pm} \rightarrow \tau^{\pm}\nu\widetilde{\chi}_{1}^{0})(\%) \end{array}$
\\ \hline
1373 &
$\begin{array}{c} 1228 \\ 1237 \end{array}$ &
$\begin{array}{c} 1176 \\ 1017 \end{array}$ &
$\begin{array}{c} 1213 \\ 1127 \end{array}$ &
$\begin{array}{c} 324 \\ 352 \end{array}$ &
$\begin{array}{c} 380 \\ 268 \end{array}$ &
$\begin{array}{c} 193 \\ 175 \end{array}$ &
$\begin{array}{c} 809 \\ 193 \end{array}$ &
$\begin{array}{c} 3.1 \\ 17.7 \\ 33.0 \\ 46.2 \end{array}$ &
$\begin{array}{c} 59.4 \\ 26.7 \\ 13.9 \end{array}$
\\ \hline \hline
\end{tabular}
\end{center}
\end{table}

\begin{table}[t]
\caption{Low energy supersymmetric particles and masses (in GeV) for $\textit{ID6BraneP1}$ point, $\Theta_{1}$ = -0.06, $\Theta_{2}$ = 0.58, $M3$ = 844, $M2$ = 351, $M1$ = 611, $m_{H}$ = 69, $m_{L}$ = 376, $m_{R}$ = 435, $A_{0}$ = -67, tan$\beta$ = 25, $m_{3/2}$ = 700. The relic density for this point is $\Omega_{\chi}$ = 0.1117. Here, $l = (e,\mu)$.}
 \label{tab:P1_WMAP_2}
\begin{center}
\begin{tabular}{c c c c c c c c c c}
\hline \hline
$\widetilde{g}$ &
$\begin{array}{c} \widetilde{u}_{L} \\ \widetilde{u}_{R} \end{array}$ &
$\begin{array}{c} \widetilde{t}_{2} \\ \widetilde{t}_{1} \end{array}$ &
$\begin{array}{c} \widetilde{b}_{2} \\ \widetilde{b}_{1} \end{array}$ &
$\begin{array}{c} \widetilde{e}_{L} \\ \widetilde{e}_{R} \end{array}$ &
$\begin{array}{c} \widetilde{\tau}_{2} \\ \widetilde{\tau}_{1} \end{array}$ &
$\begin{array}{c} \widetilde{\chi}_{2}^{0} \\ \widetilde{\chi}_{1}^{0} \end{array}$ &
$\begin{array}{c} \widetilde{\chi}_{2}^{\pm} \\ \widetilde{\chi}_{1}^{\pm} \end{array}$ &
$\begin{array}{c} Br(\widetilde{\chi}_{2}^{0} \rightarrow q\overline{q}\widetilde{\chi}_{1}^{0})(\%) \\ Br(\widetilde{\chi}_{2}^{0} \rightarrow l^{+}l^{-}\widetilde{\chi}_{1}^{0})(\%) \\ Br(\widetilde{\chi}_{2}^{0} \rightarrow \tau^{+}\tau^{-}\widetilde{\chi}_{1}^{0})(\%) \\ Br(\widetilde{\chi}_{2}^{0} \rightarrow \nu\overline{\nu}\widetilde{\chi}_{1}^{0})(\%) \end{array}$ &
$\begin{array}{c} Br(\widetilde{\chi}_{1}^{\pm} \rightarrow q\overline{q}\widetilde{\chi}_{1}^{0})(\%) \\ Br(\widetilde{\chi}_{1}^{\pm} \rightarrow l^{\pm}\nu\widetilde{\chi}_{1}^{0})(\%) \\ Br(\widetilde{\chi}_{1}^{\pm} \rightarrow \tau^{\pm}\nu\widetilde{\chi}_{1}^{0})(\%) \end{array}$
\\ \hline
1873 &
$\begin{array}{c} 1669 \\ 1683 \end{array}$ &
$\begin{array}{c} 1572 \\ 1401 \end{array}$ &
$\begin{array}{c} 1646 \\ 1534 \end{array}$ &
$\begin{array}{c} 446 \\ 490 \end{array}$ &
$\begin{array}{c} 508 \\ 396 \end{array}$ &
$\begin{array}{c} 275 \\ 254 \end{array}$ &
$\begin{array}{c} 1091 \\ 275 \end{array}$ &
$\begin{array}{c} 3.4 \\ 22.8 \\ 26.4 \\ 47.4 \end{array}$ &
$\begin{array}{c} 62.1 \\ 25.2 \\ 12.7 \end{array}$
\\ \hline \hline
\end{tabular}
\end{center}
\end{table}

\begin{table}[t]
\caption{Low energy supersymmetric particles and masses (in GeV) for $\textit{ID6BraneP1}$ point, $\Theta_{1}$ = -0.43, $\Theta_{2}$ = 0.47, $M3$ = 537, $M2$ = 203, $M1$ = 303, $m_{H}$ = 164, $m_{L}$ = 406, $m_{R}$ = 298, $A_{0}$ = 186, tan$\beta$ = 25, $m_{3/2}$ = 500. The relic density for this point is $\Omega_{\chi}$ = 1.0076. Here, $l = (e,\mu)$.}
 \label{tab:P1_SSC_1}
\begin{center}
\begin{tabular}{c c c c c c c c c c}
\hline \hline
$\widetilde{g}$ &
$\begin{array}{c} \widetilde{u}_{L} \\ \widetilde{u}_{R} \end{array}$ &
$\begin{array}{c} \widetilde{t}_{2} \\ \widetilde{t}_{1} \end{array}$ &
$\begin{array}{c} \widetilde{b}_{2} \\ \widetilde{b}_{1} \end{array}$ &
$\begin{array}{c} \widetilde{e}_{L} \\ \widetilde{e}_{R} \end{array}$ &
$\begin{array}{c} \widetilde{\tau}_{2} \\ \widetilde{\tau}_{1} \end{array}$ &
$\begin{array}{c} \widetilde{\chi}_{2}^{0} \\ \widetilde{\chi}_{1}^{0} \end{array}$ &
$\begin{array}{c} \widetilde{\chi}_{2}^{\pm} \\ \widetilde{\chi}_{1}^{\pm} \end{array}$ &
$\begin{array}{c} Br(\widetilde{\chi}_{2}^{0} \rightarrow q\overline{q}\widetilde{\chi}_{1}^{0})(\%) \\ Br(\widetilde{\chi}_{2}^{0} \rightarrow l^{+}l^{-}\widetilde{\chi}_{1}^{0})(\%) \\ Br(\widetilde{\chi}_{2}^{0} \rightarrow \tau^{+}\tau^{-}\widetilde{\chi}_{1}^{0})(\%) \\ Br(\widetilde{\chi}_{2}^{0} \rightarrow \nu\overline{\nu}\widetilde{\chi}_{1}^{0})(\%) \end{array}$ &
$\begin{array}{c} Br(\widetilde{\chi}_{1}^{\pm} \rightarrow q\overline{q}\widetilde{\chi}_{1}^{0})(\%) \\ Br(\widetilde{\chi}_{1}^{\pm} \rightarrow l^{\pm}\nu\widetilde{\chi}_{1}^{0})(\%) \\ Br(\widetilde{\chi}_{1}^{\pm} \rightarrow \tau^{\pm}\nu\widetilde{\chi}_{1}^{0})(\%) \end{array}$
\\ \hline
1239 &
$\begin{array}{c} 1149 \\ 1116 \end{array}$ &
$\begin{array}{c} 1098 \\ 929 \end{array}$ &
$\begin{array}{c} 1103 \\ 1057 \end{array}$ &
$\begin{array}{c} 428 \\ 320 \end{array}$ &
$\begin{array}{c} 433 \\ 285 \end{array}$ &
$\begin{array}{c} 154 \\ 122 \end{array}$ &
$\begin{array}{c} 722 \\ 154 \end{array}$ &
$\begin{array}{c} 37.3 \\ 21.6 \\ 10.3 \\ 30.8 \end{array}$ &
$\begin{array}{c} 64.9 \\ 23.3 \\ 11.8 \end{array}$
\\ \hline \hline
\end{tabular}
\end{center}
\end{table}

\begin{table}[t]
\caption{Low energy supersymmetric particles and masses (in GeV) for $\textit{ID6BraneP1}$ point, $\Theta_{1}$ = -0.19, $\Theta_{2}$ = 0.66, $M3$ = 600, $M2$ = 285, $M1$ = 379, $m_{H}$ = 227, $m_{L}$ = 327, $m_{R}$ = 306, $A_{0}$ = -11, tan$\beta$ = 25, $m_{3/2}$ = 500. The relic density for this point is $\Omega_{\chi}$ = 0.9166. Here, $l = (e,\mu)$.}
 \label{tab:P1_SSC_2}
\begin{center}
\begin{tabular}{c c c c c c c c c c}
\hline \hline
$\widetilde{g}$ &
$\begin{array}{c} \widetilde{u}_{L} \\ \widetilde{u}_{R} \end{array}$ &
$\begin{array}{c} \widetilde{t}_{2} \\ \widetilde{t}_{1} \end{array}$ &
$\begin{array}{c} \widetilde{b}_{2} \\ \widetilde{b}_{1} \end{array}$ &
$\begin{array}{c} \widetilde{e}_{L} \\ \widetilde{e}_{R} \end{array}$ &
$\begin{array}{c} \widetilde{\tau}_{2} \\ \widetilde{\tau}_{1} \end{array}$ &
$\begin{array}{c} \widetilde{\chi}_{2}^{0} \\ \widetilde{\chi}_{1}^{0} \end{array}$ &
$\begin{array}{c} \widetilde{\chi}_{2}^{\pm} \\ \widetilde{\chi}_{1}^{\pm} \end{array}$ &
$\begin{array}{c} Br(\widetilde{\chi}_{2}^{0} \rightarrow q\overline{q}\widetilde{\chi}_{1}^{0})(\%) \\ Br(\widetilde{\chi}_{2}^{0} \rightarrow l^{+}l^{-}\widetilde{\chi}_{1}^{0})(\%) \\ Br(\widetilde{\chi}_{2}^{0} \rightarrow \tau^{+}\tau^{-}\widetilde{\chi}_{1}^{0})(\%) \\ Br(\widetilde{\chi}_{2}^{0} \rightarrow \nu\overline{\nu}\widetilde{\chi}_{1}^{0})(\%) \end{array}$ &
$\begin{array}{c} Br(\widetilde{\chi}_{1}^{\pm} \rightarrow q\overline{q}\widetilde{\chi}_{1}^{0})(\%) \\ Br(\widetilde{\chi}_{1}^{\pm} \rightarrow l^{\pm}\nu\widetilde{\chi}_{1}^{0})(\%) \\ Br(\widetilde{\chi}_{1}^{\pm} \rightarrow \tau^{\pm}\nu\widetilde{\chi}_{1}^{0})(\%) \end{array}$
\\ \hline
1368 &
$\begin{array}{c} 1238 \\ 1229 \end{array}$ &
$\begin{array}{c} 1178 \\ 1007 \end{array}$ &
$\begin{array}{c} 1206 \\ 1134 \end{array}$ &
$\begin{array}{c} 378 \\ 338 \end{array}$ &
$\begin{array}{c} 398 \\ 286 \end{array}$ &
$\begin{array}{c} 221 \\ 154 \end{array}$ &
$\begin{array}{c} 787 \\ 221 \end{array}$ &
$\begin{array}{c} 19.7 \\ 27.8 \\ 10.0 \\ 42.5\end{array}$ &
$\begin{array}{c} 59.2 \\ 26.8 \\ 14.0 \end{array}$
\\ \hline \hline
\end{tabular}
\end{center}
\end{table}

\begin{table}[t]
\caption{Low energy supersymmetric particles and masses (in GeV) for $\textit{ID6BraneP1}$ point, $\Theta_{1}$ = 0.2, $\Theta_{2}$ = 0.69, $M3$ = 839, $M2$ = 418, $M1$ = 661, $m_{H}$ = 366, $m_{L}$ = 384, $m_{R}$ = 322, $A_{0}$ = -336, tan$\beta$ = 25, $m_{3/2}$ = 700. The relic density for this point is $\Omega_{\chi}$ = 1.0790. Here, $l = (e,\mu)$.}
 \label{tab:P1_SSC_3}
\begin{center}
\begin{tabular}{c c c c c c c c c c}
\hline \hline
$\widetilde{g}$ &
$\begin{array}{c} \widetilde{u}_{L} \\ \widetilde{u}_{R} \end{array}$ &
$\begin{array}{c} \widetilde{t}_{2} \\ \widetilde{t}_{1} \end{array}$ &
$\begin{array}{c} \widetilde{b}_{2} \\ \widetilde{b}_{1} \end{array}$ &
$\begin{array}{c} \widetilde{e}_{L} \\ \widetilde{e}_{R} \end{array}$ &
$\begin{array}{c} \widetilde{\tau}_{2} \\ \widetilde{\tau}_{1} \end{array}$ &
$\begin{array}{c} \widetilde{\chi}_{2}^{0} \\ \widetilde{\chi}_{1}^{0} \end{array}$ &
$\begin{array}{c} \widetilde{\chi}_{2}^{\pm} \\ \widetilde{\chi}_{1}^{\pm} \end{array}$ &
$\begin{array}{c} Br(\widetilde{\chi}_{2}^{0} \rightarrow q\overline{q}\widetilde{\chi}_{1}^{0})(\%) \\ Br(\widetilde{\chi}_{2}^{0} \rightarrow l^{+}l^{-}\widetilde{\chi}_{1}^{0})(\%) \\ Br(\widetilde{\chi}_{2}^{0} \rightarrow \tau^{+}\tau^{-}\widetilde{\chi}_{1}^{0})(\%) \\ Br(\widetilde{\chi}_{2}^{0} \rightarrow \nu\overline{\nu}\widetilde{\chi}_{1}^{0})(\%) \end{array}$ &
$\begin{array}{c} Br(\widetilde{\chi}_{1}^{\pm} \rightarrow q\overline{q}\widetilde{\chi}_{1}^{0})(\%) \\ Br(\widetilde{\chi}_{1}^{\pm} \rightarrow l^{\pm}\nu\widetilde{\chi}_{1}^{0})(\%) \\ Br(\widetilde{\chi}_{1}^{\pm} \rightarrow \tau^{\pm}\nu\widetilde{\chi}_{1}^{0})(\%) \end{array}$
\\ \hline
1862 &
$\begin{array}{c} 1670 \\ 1651 \end{array}$ &
$\begin{array}{c} 1554 \\ 1340 \end{array}$ &
$\begin{array}{c} 1607 \\ 1519 \end{array}$ &
$\begin{array}{c} 477 \\ 404 \end{array}$ &
$\begin{array}{c} 487 \\ 339 \end{array}$ &
$\begin{array}{c} 331 \\ 276 \end{array}$ &
$\begin{array}{c} 1072 \\ 331 \end{array}$ &
$\begin{array}{c} 6.6 \\ 28.8 \\ 14.1 \\ 50.5 \end{array}$ &
$\begin{array}{c} 57.0 \\ 27.9 \\ 15.1 \end{array}$
\\ \hline \hline
\end{tabular}
\end{center}
\end{table}

Next we study points within regions of the allowed parameter space with pattern $\textit{ID6BraneP4}$ that generate the observed WMAP dark matter density. Here, the mass difference between the $\widetilde{\tau}_{1}$ and $\widetilde{\chi}_{1}^{0}$ is nearly degenerate, so these points lie within the stau-neutralino coannihilation region of the parameter space. Three typical points from regions of the parameter space with pattern $\textit{ID6BraneP4}$ are shown in Tables~\ref{tab:P4_WMAP_1},~\ref{tab:P4_WMAP_2}, and~\ref{tab:P4_WMAP_3}. The processes with the largest production cross-sections for these points are $q+\overline{q}\rightarrow\widetilde{\chi}_{1}^{\pm}+\widetilde{\chi}_{1}^{\pm}$, $q+\overline{q}\rightarrow\widetilde{\chi}_{2}^{0}+\widetilde{\chi}_{1}^{\pm}$, and $q+\overline{q}\rightarrow\widetilde{q}+\widetilde{q}$. The $\widetilde{\chi}_{1}^{\pm}$ and $\widetilde{\chi}_{2}^{0}$ are virtually degenerate, whereas the $\widetilde{\tau}_{1}^{\pm}$ is lighter than both the $\widetilde{\chi}_{1}^{\pm}$ and $\widetilde{\chi}_{2}^{0}$, so both the $\widetilde{\chi}_{1}^{\pm}$ and $\widetilde{\chi}_{2}^{0}$ will decay to $\widetilde{\tau}_{1}^{\pm}$ nearly 100\% of the time. The second lightest neutralino will decay to tau through the decay chain $\widetilde{\chi}_{2}^{0}\rightarrow \widetilde{\tau}_{1}^{\pm}\tau^{\mp}\rightarrow\tau^{\pm}\tau^{\mp}\widetilde{\chi}_{1}^{0}$, while the chargino will also decay to tau through $\widetilde{\chi}_{1}^{\pm}\rightarrow \widetilde{\tau}_{1}^{\pm}\nu\rightarrow\tau^{\pm}\nu\widetilde{\chi}_{1}^{0}$, so the results of both decay chains will be low energy tau. The squark decay chain will provide jets through $\widetilde{q}_{R} \rightarrow q\widetilde{\chi}_{1}^{0}$, $\widetilde{q}_{L} \rightarrow q\widetilde{\chi}_{2}^{0}$, and $\widetilde{q}_{L} \rightarrow q\widetilde{\chi}_{1}^{\pm}$. As the mass parameters were varied, there was no change in the final states. Therefore, we expect to see the signal $\tau + jets + \mbox{E\!\!\!\!/}_{T}$ at LHC for these points that produce the WMAP observed dark matter density.

We now analyze points within regions of the parameter space with pattern $\textit{ID6BraneP4}$ that generate the diluted dark matter density in the context of SSC. The three sample points from regions of the parameter space with pattern $\textit{ID6BraneP4}$ are shown in Tables~\ref{tab:P4_SSC_1},~\ref{tab:P4_SSC_2}, and~\ref{tab:P4_SSC_3}. Here the mass difference between the $\widetilde{\tau}_{1}$ and $\widetilde{\chi}_{1}^{0}$ is much greater than those points in the WMAP region, so we are no longer within the stau-neutralino coannihilation region, so accordingly, the relic density is larger. The $\widetilde{\chi}_{1}^{\pm}$ and $\widetilde{\chi}_{2}^{0}$ are still virtually degenerate, so both will decay to $\widetilde{\tau}_{1}$ almost 100\% of the time, with the exception of the point in Table~\ref{tab:P4_SSC_2}. The mass parameters for this point will produce some $W^{\pm}$ bosons through $\widetilde{\chi}_{1}^{\pm} \rightarrow W^{\pm}\widetilde{\chi}_{1}^{0}$ with a branching ratio of 25.5\%, though $\widetilde{\chi}_{1}^{\pm} \rightarrow \widetilde{\tau}^{\pm}\nu$ is still the dominant decay mode. Otherwise, the dominant decay chains and source of the jets remain the same as the WMAP regions, as well as the processes with the largest cross-sections. In essence, we expect the same signals as those of the WMAP regions, but the signals in the SSC regions should be clearly discriminated from the WMAP regions by observing a larger mass difference between the $\widetilde{\tau}_{1}$ with the $\widetilde{\chi}_{1}^{0}$, namely the production of high energy tau as opposed to the low energy tau in the WMAP region.

\begin{table}[t]
\caption{Low energy supersymmetric particles and masses (in GeV) for $\textit{ID6BraneP4}$ point, $\Theta_{1}$ = -0.19, $\Theta_{2}$ = 0.75, $M3$ = 599, $M2$ = 324, $M1$ = 354, $m_{H}$ = 315, $m_{L}$ = 346, $m_{R}$ = 292, $A_{0}$ = -50, tan$\beta$ = 46, $m_{3/2}$ = 500. The relic density for this point is $\Omega_{\chi}$ = 0.1166.}
 \label{tab:P4_WMAP_1}
\begin{center}
\begin{tabular}{c c c c c c c c c}
\hline \hline
$\widetilde{g}$ &
$\begin{array}{c} \widetilde{u}_{L} \\ \widetilde{u}_{R} \end{array}$ &
$\begin{array}{c} \widetilde{t}_{2} \\ \widetilde{t}_{1} \end{array}$ &
$\begin{array}{c} \widetilde{b}_{2} \\ \widetilde{b}_{1} \end{array}$ &
$\begin{array}{c} \widetilde{e}_{L} \\ \widetilde{e}_{R} \end{array}$ &
$\begin{array}{c} \widetilde{\tau}_{2} \\ \widetilde{\tau}_{1} \end{array}$ &
$\begin{array}{c} \widetilde{\chi}_{2}^{0} \\ \widetilde{\chi}_{1}^{0} \end{array}$ &
$\begin{array}{c} \widetilde{\chi}_{2}^{\pm} \\ \widetilde{\chi}_{1}^{\pm} \end{array}$ &
$\begin{array}{c} Br(\widetilde{\chi}_{2}^{0} \rightarrow \widetilde{\tau}_{1}\tau)(\%) \\ Br(\widetilde{\tau}_{1} \rightarrow \tau\widetilde{\chi}_{1}^{0})(\%) \\ Br(\widetilde{\chi}_{1}^{\pm} \rightarrow \widetilde{\tau}^{\pm}\nu)(\%) \end{array}$
\\ \hline
1363 &
$\begin{array}{c} 1244 \\ 1222 \end{array}$ &
$\begin{array}{c} 1150 \\ 989 \end{array}$ &
$\begin{array}{c} 1157 \\ 1087 \end{array}$ &
$\begin{array}{c} 406 \\ 321 \end{array}$ &
$\begin{array}{c} 422 \\ 161 \end{array}$ &
$\begin{array}{c} 254 \\ 144 \end{array}$ &
$\begin{array}{c} 760 \\ 254 \end{array}$ &
$\begin{array}{c} 99.9 \\ 100.0 \\ 99.6 \end{array}$
\\ \hline \hline
\end{tabular}
\end{center}
\end{table}

\begin{table}[t]
\caption{Low energy supersymmetric particles and masses (in GeV) for $\textit{ID6BraneP4}$ point, $\Theta_{1}$ = -0.59, $\Theta_{2}$ = 0.46, $M3$ = 681, $M2$ = 278, $M1$ = 299, $m_{H}$ = 407, $m_{L}$ = 693, $m_{R}$ = 349, $A_{0}$ = 371, tan$\beta$ = 46, $m_{3/2}$ = 700. The relic density for this point is $\Omega_{\chi}$ = 0.1130.}
 \label{tab:P4_WMAP_2}
\begin{center}
\begin{tabular}{c c c c c c c c c}
\hline \hline
$\widetilde{g}$ &
$\begin{array}{c} \widetilde{u}_{L} \\ \widetilde{u}_{R} \end{array}$ &
$\begin{array}{c} \widetilde{t}_{2} \\ \widetilde{t}_{1} \end{array}$ &
$\begin{array}{c} \widetilde{b}_{2} \\ \widetilde{b}_{1} \end{array}$ &
$\begin{array}{c} \widetilde{e}_{L} \\ \widetilde{e}_{R} \end{array}$ &
$\begin{array}{c} \widetilde{\tau}_{2} \\ \widetilde{\tau}_{1} \end{array}$ &
$\begin{array}{c} \widetilde{\chi}_{2}^{0} \\ \widetilde{\chi}_{1}^{0} \end{array}$ &
$\begin{array}{c} \widetilde{\chi}_{2}^{\pm} \\ \widetilde{\chi}_{1}^{\pm} \end{array}$ &
$\begin{array}{c} Br(\widetilde{\chi}_{2}^{0} \rightarrow \widetilde{\tau}_{1}\tau)(\%) \\ Br(\widetilde{\tau}_{1} \rightarrow \tau\widetilde{\chi}_{1}^{0})(\%) \\ Br(\widetilde{\chi}_{1}^{\pm} \rightarrow \widetilde{\tau}^{\pm}\nu)(\%) \end{array}$
\\ \hline
1545 &
$\begin{array}{c} 1501 \\ 1374 \end{array}$ &
$\begin{array}{c} 1376 \\ 1135 \end{array}$ &
$\begin{array}{c} 1374 \\ 1267 \end{array}$ &
$\begin{array}{c} 713 \\ 365 \end{array}$ &
$\begin{array}{c} 684 \\ 132 \end{array}$ &
$\begin{array}{c} 217 \\ 121 \end{array}$ &
$\begin{array}{c} 859 \\ 217 \end{array}$ &
$\begin{array}{c} 99.9 \\ 100.0 \\ 99.5 \end{array}$
\\ \hline \hline
\end{tabular}
\end{center}
\end{table}

\begin{table}[t]
\caption{Low energy supersymmetric particles and masses (in GeV) for $\textit{ID6BraneP4}$ point, $\Theta_{1}$ = 0.02, $\Theta_{2}$ = 0.68, $M3$ = 856, $M2$ = 412, $M1$ = 616, $m_{H}$ = 308, $m_{L}$ = 381, $m_{R}$ = 396, $A_{0}$ = -186, tan$\beta$ = 46, $m_{3/2}$ = 700. The relic density for this point is $\Omega_{\chi}$ = 0.1128.}
 \label{tab:P4_WMAP_3}
\begin{center}
\begin{tabular}{c c c c c c c c c}
\hline \hline
$\widetilde{g}$ &
$\begin{array}{c} \widetilde{u}_{L} \\ \widetilde{u}_{R} \end{array}$ &
$\begin{array}{c} \widetilde{t}_{2} \\ \widetilde{t}_{1} \end{array}$ &
$\begin{array}{c} \widetilde{b}_{2} \\ \widetilde{b}_{1} \end{array}$ &
$\begin{array}{c} \widetilde{e}_{L} \\ \widetilde{e}_{R} \end{array}$ &
$\begin{array}{c} \widetilde{\tau}_{2} \\ \widetilde{\tau}_{1} \end{array}$ &
$\begin{array}{c} \widetilde{\chi}_{2}^{0} \\ \widetilde{\chi}_{1}^{0} \end{array}$ &
$\begin{array}{c} \widetilde{\chi}_{2}^{\pm} \\ \widetilde{\chi}_{1}^{\pm} \end{array}$ &
$\begin{array}{c} Br(\widetilde{\chi}_{2}^{0} \rightarrow \widetilde{\tau}_{1}\tau)(\%) \\ Br(\widetilde{\tau}_{1} \rightarrow \tau\widetilde{\chi}_{1}^{0})(\%) \\ Br(\widetilde{\chi}_{1}^{\pm} \rightarrow \widetilde{\tau}^{\pm}\nu)(\%) \end{array}$
\\ \hline
1895 &
$\begin{array}{c} 1694 \\ 1692 \end{array}$ &
$\begin{array}{c} 1551 \\ 1389 \end{array}$ &
$\begin{array}{c} 1586 \\ 1498 \end{array}$ &
$\begin{array}{c} 471 \\ 457 \end{array}$ &
$\begin{array}{c} 512 \\ 276 \end{array}$ &
$\begin{array}{c} 327 \\ 257 \end{array}$ &
$\begin{array}{c} 1073 \\ 327 \end{array}$ &
$\begin{array}{c} 99.9 \\ 100.0 \\ 99.9\end{array}$
\\ \hline \hline
\end{tabular}
\end{center}
\end{table}

\begin{table}[t]
\caption{Low energy supersymmetric particles and masses (in GeV) for $\textit{ID6BraneP4}$ point, $\Theta_{1}$ = -0.44, $\Theta_{2}$ = 0.58, $M3$ = 548, $M2$ = 251, $M1$ = 283, $m_{H}$ = 271, $m_{L}$ = 431, $m_{R}$ = 291, $A_{0}$ = 147, tan$\beta$ = 46, $m_{3/2}$ = 500. The relic density for this point is $\Omega_{\chi}$ = 0.5003.}
 \label{tab:P4_SSC_1}
\begin{center}
\begin{tabular}{c c c c c c c c c}
\hline \hline
$\widetilde{g}$ &
$\begin{array}{c} \widetilde{u}_{L} \\ \widetilde{u}_{R} \end{array}$ &
$\begin{array}{c} \widetilde{t}_{2} \\ \widetilde{t}_{1} \end{array}$ &
$\begin{array}{c} \widetilde{b}_{2} \\ \widetilde{b}_{1} \end{array}$ &
$\begin{array}{c} \widetilde{e}_{L} \\ \widetilde{e}_{R} \end{array}$ &
$\begin{array}{c} \widetilde{\tau}_{2} \\ \widetilde{\tau}_{1} \end{array}$ &
$\begin{array}{c} \widetilde{\chi}_{2}^{0} \\ \widetilde{\chi}_{1}^{0} \end{array}$ &
$\begin{array}{c} \widetilde{\chi}_{2}^{\pm} \\ \widetilde{\chi}_{1}^{\pm} \end{array}$ &
$\begin{array}{c} Br(\widetilde{\chi}_{2}^{0} \rightarrow \widetilde{\tau}_{1}\tau)(\%) \\ Br(\widetilde{\tau}_{1} \rightarrow \tau\widetilde{\chi}_{1}^{0})(\%) \\ Br(\widetilde{\chi}_{1}^{\pm} \rightarrow \widetilde{\tau}^{\pm}\nu)(\%) \end{array}$
\\ \hline
1260 &
$\begin{array}{c} 1178 \\ 1131 \end{array}$ &
$\begin{array}{c} 1095 \\ 930 \end{array}$ &
$\begin{array}{c} 1094 \\ 1026 \end{array}$ &
$\begin{array}{c} 461 \\ 310 \end{array}$ &
$\begin{array}{c} 460 \\ 173 \end{array}$ &
$\begin{array}{c} 195 \\ 114 \end{array}$ &
$\begin{array}{c} 711 \\ 195 \end{array}$ &
$\begin{array}{c} 99.9 \\ 100.0 \\ 99.9 \end{array}$
\\ \hline \hline
\end{tabular}
\end{center}
\end{table}

\begin{table}[t]
\caption{Low energy supersymmetric particles and masses (in GeV) for $\textit{ID6BraneP4}$ point, $\Theta_{1}$ = -0.51, $\Theta_{2}$ = 0.52, $M3$ = 730, $M2$ = 315, $M1$ = 355, $m_{H}$ = 380, $m_{L}$ = 642, $m_{R}$ = 387, $A_{0}$ = 286, tan$\beta$ = 46, $m_{3/2}$ = 700. The relic density for this point is $\Omega_{\chi}$ = 1.0030.}
 \label{tab:P4_SSC_2}
\begin{center}
\begin{tabular}{c c c c c c c c c}
\hline \hline
$\widetilde{g}$ &
$\begin{array}{c} \widetilde{u}_{L} \\ \widetilde{u}_{R} \end{array}$ &
$\begin{array}{c} \widetilde{t}_{2} \\ \widetilde{t}_{1} \end{array}$ &
$\begin{array}{c} \widetilde{b}_{2} \\ \widetilde{b}_{1} \end{array}$ &
$\begin{array}{c} \widetilde{e}_{L} \\ \widetilde{e}_{R} \end{array}$ &
$\begin{array}{c} \widetilde{\tau}_{2} \\ \widetilde{\tau}_{1} \end{array}$ &
$\begin{array}{c} \widetilde{\chi}_{2}^{0} \\ \widetilde{\chi}_{1}^{0} \end{array}$ &
$\begin{array}{c} \widetilde{\chi}_{2}^{\pm} \\ \widetilde{\chi}_{1}^{\pm} \end{array}$ &
$\begin{array}{c} Br(\widetilde{\chi}_{2}^{0} \rightarrow \widetilde{\tau}_{1}\tau)(\%) \\ Br(\widetilde{\tau}_{1} \rightarrow \tau\widetilde{\chi}_{1}^{0})(\%) \\ Br(\widetilde{\chi}_{1}^{\pm} \rightarrow \widetilde{\tau}^{\pm}\nu)(\%) \end{array}$
\\ \hline
1644 &
$\begin{array}{c} 1560 \\ 1469 \end{array}$ &
$\begin{array}{c} 1430 \\ 1220 \end{array}$ &
$\begin{array}{c} 1431 \\ 1352 \end{array}$ &
$\begin{array}{c} 672 \\ 408 \end{array}$ &
$\begin{array}{c} 649 \\ 238 \end{array}$ &
$\begin{array}{c} 247 \\ 145 \end{array}$ &
$\begin{array}{c} 917 \\ 247 \end{array}$ &
$\begin{array}{c} 94.6 \\ 100.0 \\ 74.5 \end{array}$
\\ \hline \hline
\end{tabular}
\end{center}
\end{table}

\begin{table}[t]
\caption{Low energy supersymmetric particles and masses (in GeV) for $\textit{ID6BraneP4}$ point, $\Theta_{1}$ = -0.27, $\Theta_{2}$ = 0.76, $M3$ = 819, $M2$ = 460, $M1$ = 444, $m_{H}$ = 482, $m_{L}$ = 535, $m_{R}$ = 408, $A_{0}$ = -20, tan$\beta$ = 46, $m_{3/2}$ = 700. The relic density for this point is $\Omega_{\chi}$ = 1.0521.}
 \label{tab:P4_SSC_3}
\begin{center}
\begin{tabular}{c c c c c c c c c}
\hline \hline
$\widetilde{g}$ &
$\begin{array}{c} \widetilde{u}_{L} \\ \widetilde{u}_{R} \end{array}$ &
$\begin{array}{c} \widetilde{t}_{2} \\ \widetilde{t}_{1} \end{array}$ &
$\begin{array}{c} \widetilde{b}_{2} \\ \widetilde{b}_{1} \end{array}$ &
$\begin{array}{c} \widetilde{e}_{L} \\ \widetilde{e}_{R} \end{array}$ &
$\begin{array}{c} \widetilde{\tau}_{2} \\ \widetilde{\tau}_{1} \end{array}$ &
$\begin{array}{c} \widetilde{\chi}_{2}^{0} \\ \widetilde{\chi}_{1}^{0} \end{array}$ &
$\begin{array}{c} \widetilde{\chi}_{2}^{\pm} \\ \widetilde{\chi}_{1}^{\pm} \end{array}$ &
$\begin{array}{c} Br(\widetilde{\chi}_{2}^{0} \rightarrow \widetilde{\tau}_{1}\tau)(\%) \\ Br(\widetilde{\tau}_{1} \rightarrow \tau\widetilde{\chi}_{1}^{0})(\%) \\ Br(\widetilde{\chi}_{1}^{\pm} \rightarrow \widetilde{\tau}^{\pm}\nu)(\%) \end{array}$
\\ \hline
1821 &
$\begin{array}{c} 1677 \\ 1627 \end{array}$ &
$\begin{array}{c} 1523 \\ 1331 \end{array}$ &
$\begin{array}{c} 1535 \\ 1469 \end{array}$ &
$\begin{array}{c} 611 \\ 439 \end{array}$ &
$\begin{array}{c} 594 \\ 267 \end{array}$ &
$\begin{array}{c} 367 \\ 184 \end{array}$ &
$\begin{array}{c} 986 \\ 367 \end{array}$ &
$\begin{array}{c} 98.1 \\ 100.0 \\ 98.0 \end{array}$
\\ \hline \hline
\end{tabular}
\end{center}
\end{table}

Lastly, we consider points within regions of the allowed parameter space with pattern $\textit{ID6BraneP5}$ that generate the observed WMAP dark matter density. In this region, the mass difference between the $\widetilde{\tau}_{1}$ and $\widetilde{\chi}_{1}^{0}$ is nearly degenerate, so these points lie within the stau-neutralino coannihilation region. Two representative points from regions of the parameter space with pattern $\textit{ID6BraneP5}$ are shown in Table~\ref{tab:P5_WMAP_1} and Table~\ref{tab:P5_WMAP_2}. The processes with the largest production cross-sections for these points are $q+\overline{q}\rightarrow\widetilde{q}+\widetilde{q}$, $q+\overline{q}\rightarrow\widetilde{q}+\widetilde{g}$, $q+\overline{q}\rightarrow\widetilde{\chi}_{1}^{\pm}+\widetilde{\chi}_{1}^{\pm}$, and $q+\overline{q}\rightarrow\widetilde{\chi}_{2}^{0}+\widetilde{\chi}_{1}^{\pm}$. The $\widetilde{\chi}_{1}^{\pm}$ and $\widetilde{\chi}_{2}^{0}$ are virtually degenerate, and the $\widetilde{\tau}_{1}^{\pm}$ is lighter than both the $\widetilde{\chi}_{1}^{\pm}$ and $\widetilde{\chi}_{2}^{0}$, so  the decay chains for the $\widetilde{\chi}_{1}^{\pm}$ and $\widetilde{\chi}_{2}^{0}$ are the same as those for regions of the parameter space with pattern $\textit{ID6BraneP4}$ that generate the WMAP observed relic density, resulting in low energy tau. However, with a larger gluino production cross-section, we can include the process $\widetilde{g} \rightarrow q\widetilde{q}$ as one of the primary sources of jets, in addition to the squark decay chains $\widetilde{q}_{R} \rightarrow q\widetilde{\chi}_{1}^{0}$, $\widetilde{q}_{L} \rightarrow q\widetilde{\chi}_{2}^{0}$, and $\widetilde{q}_{L} \rightarrow q\widetilde{\chi}_{1}^{\pm}$. Thus, we anticipate the signal $\tau + jets + \mbox{E\!\!\!\!/}_{T}$ at LHC for these points that produce the WMAP observed dark matter density.

To conclude the discussion of the final states, we investigate points within regions of the parameter space with pattern $\textit{ID6BraneP5}$ that generate the diluted dark matter density in the SSC scenario. Two representative points from regions of the parameter space with pattern $\textit{ID6BraneP5}$ are shown in Table~\ref{tab:P5_SSC_1} and Table~\ref{tab:P5_SSC_2}. Here the mass difference between the $\widetilde{\tau}_{1}$ and $\widetilde{\chi}_{1}^{0}$ is much larger than those points in the WMAP region, thus, these points do not reside within the stau-neutralino coannihilation region. The $\widetilde{\chi}_{1}^{\pm}$ and $\widetilde{\chi}_{2}^{0}$ are still virtually degenerate, so the dominant decay mode for both is to $\widetilde{\tau}_{1}$, but not necessarily 100\% of the time. We do find for the point in Table~\ref{tab:P5_SSC_1} a small branching ratio of 16.2\% for the production of the lightest Higgs boson through $\widetilde{\chi}_{2}^{0} \rightarrow h_{0}\widetilde{\chi}_{1}^{0}$, though the dominant decay chain remains $\widetilde{\chi}_{2}^{0} \rightarrow \widetilde{\tau}_{1}\tau$.  The lower mass parameters for the point in Table~\ref{tab:P5_SSC_1} will also produce some $W^{\pm}$ bosons through $\widetilde{\chi}_{1}^{\pm} \rightarrow W^{\pm}\widetilde{\chi}_{1}^{0}$ with a branching ratio of 21.7\%, though $\widetilde{\chi}_{1}^{\pm} \rightarrow \widetilde{\tau}^{\pm}\nu$ is still the dominant decay mode. The $h_{0}$ and $W^{\pm}$ branching ratios decrease as the mass parameters are increased. Other than these differences, the dominant decay chains and source of the jets remain the same as the WMAP regions, as well as the processes with the largest production cross-sections. Thus, we foresee similar signals as those of the WMAP regions, however, the signals in the SSC regions should be distinguished from the WMAP regions by observation of the much larger mass difference between the $\widetilde{\tau}_{1}$ with the $\widetilde{\chi}_{1}^{0}$, resulting in high energy tau, in contrast to low energy tau in the WMAP region.

\begin{table}[t]
\caption{Low energy supersymmetric particles and masses (in GeV) for $\textit{ID6BraneP5}$ point, $\Theta_{1}$ = 0.14, $\Theta_{2}$ = 0.88, $M3$ = 577, $M2$ = 381, $M1$ = 385, $m_{H}$ = 415, $m_{L}$ = 305, $m_{R}$ = 187, $A_{0}$ = -295, tan$\beta$ = 25, $m_{3/2}$ = 500. The relic density for this point is $\Omega_{\chi}$ = 0.1118.}
 \label{tab:P5_WMAP_1}
\begin{center}
\begin{tabular}{c c c c c c c c c c}
\hline \hline
$\widetilde{g}$ &
$\begin{array}{c} \widetilde{u}_{L} \\ \widetilde{u}_{R} \end{array}$ &
$\begin{array}{c} \widetilde{t}_{2} \\ \widetilde{t}_{1} \end{array}$ &
$\begin{array}{c} \widetilde{b}_{2} \\ \widetilde{b}_{1} \end{array}$ &
$\begin{array}{c} \widetilde{e}_{L} \\ \widetilde{e}_{R} \end{array}$ &
$\begin{array}{c} \widetilde{\tau}_{2} \\ \widetilde{\tau}_{1} \end{array}$ &
$\begin{array}{c} \widetilde{\chi}_{2}^{0} \\ \widetilde{\chi}_{1}^{0} \end{array}$ &
$\begin{array}{c} \widetilde{\chi}_{2}^{\pm} \\ \widetilde{\chi}_{1}^{\pm} \end{array}$ &
$\begin{array}{c} Br(\widetilde{\chi}_{2}^{0} \rightarrow \widetilde{\tau}_{1}\tau)(\%) \\ Br(\widetilde{\tau}_{1} \rightarrow \tau\widetilde{\chi}_{1}^{0})(\%) \\ Br(\widetilde{\chi}_{1}^{\pm} \rightarrow \widetilde{\tau}^{\pm}\nu)(\%) \end{array}$
\\ \hline
1313 &
$\begin{array}{c} 1201 \\ 1161 \end{array}$ &
$\begin{array}{c} 1123 \\ 904 \end{array}$ &
$\begin{array}{c} 1131 \\ 1078 \end{array}$ &
$\begin{array}{c} 395 \\ 237 \end{array}$ &
$\begin{array}{c} 394 \\ 166 \end{array}$ &
$\begin{array}{c} 299 \\ 158 \end{array}$ &
$\begin{array}{c} 728 \\ 299 \end{array}$ &
$\begin{array}{c} 97.0 \\ 100.0 \\ 96.9 \end{array}$
\\ \hline \hline
\end{tabular}
\end{center}
\end{table}

\begin{table}[t]
\caption{Low energy supersymmetric particles and masses (in GeV) for $\textit{ID6BraneP5}$ point, $\Theta_{1}$ = 0.27, $\Theta_{2}$ = 0.73, $M3$ = 823, $M2$ = 442, $M1$ = 655, $m_{H}$ = 447, $m_{L}$ = 411, $m_{R}$ = 253, $A_{0}$ = -420, tan$\beta$ = 25, $m_{3/2}$ = 700. The relic density for this point is $\Omega_{\chi}$ = 0.1117.}
 \label{tab:P5_WMAP_2}
\begin{center}
\begin{tabular}{c c c c c c c c c c}
\hline \hline
$\widetilde{g}$ &
$\begin{array}{c} \widetilde{u}_{L} \\ \widetilde{u}_{R} \end{array}$ &
$\begin{array}{c} \widetilde{t}_{2} \\ \widetilde{t}_{1} \end{array}$ &
$\begin{array}{c} \widetilde{b}_{2} \\ \widetilde{b}_{1} \end{array}$ &
$\begin{array}{c} \widetilde{e}_{L} \\ \widetilde{e}_{R} \end{array}$ &
$\begin{array}{c} \widetilde{\tau}_{2} \\ \widetilde{\tau}_{1} \end{array}$ &
$\begin{array}{c} \widetilde{\chi}_{2}^{0} \\ \widetilde{\chi}_{1}^{0} \end{array}$ &
$\begin{array}{c} \widetilde{\chi}_{2}^{\pm} \\ \widetilde{\chi}_{1}^{\pm} \end{array}$ &
$\begin{array}{c} Br(\widetilde{\chi}_{2}^{0} \rightarrow \widetilde{\tau}_{1}\tau)(\%) \\ Br(\widetilde{\tau}_{1} \rightarrow \tau\widetilde{\chi}_{1}^{0})(\%) \\ Br(\widetilde{\chi}_{1}^{\pm} \rightarrow \widetilde{\tau}^{\pm}\nu)(\%) \end{array}$
\\ \hline
1827 &
$\begin{array}{c} 1651 \\ 1610 \end{array}$ &
$\begin{array}{c} 1528 \\ 1289 \end{array}$ &
$\begin{array}{c} 1565 \\ 1494 \end{array}$ &
$\begin{array}{c} 508 \\ 350 \end{array}$ &
$\begin{array}{c} 505 \\ 280 \end{array}$ &
$\begin{array}{c} 351 \\ 274 \end{array}$ &
$\begin{array}{c} 1043 \\ 351 \end{array}$ &
$\begin{array}{c} 99.9 \\ 100.0 \\ 99.9 \end{array}$
\\ \hline \hline
\end{tabular}
\end{center}
\end{table}

\begin{table}[t]
\caption{Low energy supersymmetric particles and masses (in GeV) for $\textit{ID6BraneP5}$ point, $\Theta_{1}$ = -0.23, $\Theta_{2}$ = 0.87, $M3$ = 565, $M2$ = 376, $M1$ = 279, $m_{H}$ = 422, $m_{L}$ = 389, $m_{R}$ = 271, $A_{0}$ = -89, tan$\beta$ = 25, $m_{3/2}$ = 500. The relic density for this point is $\Omega_{\chi}$ = 0.8199.}
 \label{tab:P5_SSC_1}
\begin{center}
\begin{tabular}{c c c c c c c c c c}
\hline \hline
$\widetilde{g}$ &
$\begin{array}{c} \widetilde{u}_{L} \\ \widetilde{u}_{R} \end{array}$ &
$\begin{array}{c} \widetilde{t}_{2} \\ \widetilde{t}_{1} \end{array}$ &
$\begin{array}{c} \widetilde{b}_{2} \\ \widetilde{b}_{1} \end{array}$ &
$\begin{array}{c} \widetilde{e}_{L} \\ \widetilde{e}_{R} \end{array}$ &
$\begin{array}{c} \widetilde{\tau}_{2} \\ \widetilde{\tau}_{1} \end{array}$ &
$\begin{array}{c} \widetilde{\chi}_{2}^{0} \\ \widetilde{\chi}_{1}^{0} \end{array}$ &
$\begin{array}{c} \widetilde{\chi}_{2}^{\pm} \\ \widetilde{\chi}_{1}^{\pm} \end{array}$ &
$\begin{array}{c} Br(\widetilde{\chi}_{2}^{0} \rightarrow \widetilde{\tau}_{1}\tau)(\%) \\ Br(\widetilde{\chi}_{2}^{0} \rightarrow h_{0}\widetilde{\chi}_{1}^{0})(\%) \\ Br(\widetilde{\tau}_{1} \rightarrow \tau\widetilde{\chi}_{1}^{0})(\%) \\ Br(\widetilde{\chi}_{1}^{\pm} \rightarrow \widetilde{\tau}^{\pm}\nu)(\%) \end{array}$
\\ \hline
1294 &
$\begin{array}{c} 1205 \\ 1156 \end{array}$ &
$\begin{array}{c} 1130 \\ 917 \end{array}$ &
$\begin{array}{c} 1133 \\ 1089 \end{array}$ &
$\begin{array}{c} 458 \\ 292 \end{array}$ &
$\begin{array}{c} 456 \\ 244 \end{array}$ &
$\begin{array}{c} 295 \\ 113 \end{array}$ &
$\begin{array}{c} 701 \\ 295 \end{array}$ &
$\begin{array}{c} 80.2 \\ 16.2 \\ 100.0 \\ 78.3 \end{array}$
\\ \hline \hline
\end{tabular}
\end{center}
\end{table}

\begin{table}[t]
\caption{Low energy supersymmetric particles and masses (in GeV) for $\textit{ID6BraneP5}$ point, $\Theta_{1}$ = 0.09, $\Theta_{2}$ = 0.84, $M3$ = 833, $M2$ = 509, $M1$ = 560, $m_{H}$ = 528, $m_{L}$ = 418, $m_{R}$ = 312, $A_{0}$ = -344, tan$\beta$ = 25, $m_{3/2}$ = 700. The relic density for this point is $\Omega_{\chi}$ = 1.0380.}
 \label{tab:P5_SSC_2}
\begin{center}
\begin{tabular}{c c c c c c c c c c}
\hline \hline
$\widetilde{g}$ &
$\begin{array}{c} \widetilde{u}_{L} \\ \widetilde{u}_{R} \end{array}$ &
$\begin{array}{c} \widetilde{t}_{2} \\ \widetilde{t}_{1} \end{array}$ &
$\begin{array}{c} \widetilde{b}_{2} \\ \widetilde{b}_{1} \end{array}$ &
$\begin{array}{c} \widetilde{e}_{L} \\ \widetilde{e}_{R} \end{array}$ &
$\begin{array}{c} \widetilde{\tau}_{2} \\ \widetilde{\tau}_{1} \end{array}$ &
$\begin{array}{c} \widetilde{\chi}_{2}^{0} \\ \widetilde{\chi}_{1}^{0} \end{array}$ &
$\begin{array}{c} \widetilde{\chi}_{2}^{\pm} \\ \widetilde{\chi}_{1}^{\pm} \end{array}$ &
$\begin{array}{c} Br(\widetilde{\chi}_{2}^{0} \rightarrow \widetilde{\tau}_{1}\tau)(\%) \\ Br(\widetilde{\chi}_{2}^{0} \rightarrow h_{0}\widetilde{\chi}_{1}^{0})(\%) \\ Br(\widetilde{\tau}_{1} \rightarrow \tau\widetilde{\chi}_{1}^{0})(\%) \\ Br(\widetilde{\chi}_{1}^{\pm} \rightarrow \widetilde{\tau}^{\pm}\nu)(\%) \end{array}$
\\ \hline
1847 &
$\begin{array}{c} 1674 \\ 1633 \end{array}$ &
$\begin{array}{c} 1547 \\ 1306 \end{array}$ &
$\begin{array}{c} 1588 \\ 1515 \end{array}$ &
$\begin{array}{c} 534 \\ 375 \end{array}$ &
$\begin{array}{c} 529 \\ 308 \end{array}$ &
$\begin{array}{c} 405 \\ 233 \end{array}$ &
$\begin{array}{c} 1017 \\ 405 \end{array}$ &
$\begin{array}{c} 94.6 \\ 4.3 \\ 100.0 \\ 94.5 \end{array}$
\\ \hline \hline
\end{tabular}
\end{center}
\end{table}

We now have the complete set of final states for the model in hand, so we can compare them to the expected final states for mSUGRA. In the region of the mSUGRA allowed parameter space that can generate the WMAP observed dark matter density, primarily squarks and gluinos will be produced in the stau-neutralino coannihilation region. The characteristic decay chain in this region of mSUGRA is $\widetilde{q} \rightarrow q\widetilde{\chi}_{2}^{0} \rightarrow q\tau^{\pm}\widetilde{\tau}_{1}^{\mp} \rightarrow q\tau^{\pm}\tau^{\mp}\widetilde{\chi}_{1}^{0}$. In the mSUGRA region of the allowed parameter space that can generate the relic density in the SSC scenario, the three characteristic decays are $\widetilde{\chi}_{2}^{0} \rightarrow \widetilde{\tau}_{1}^{\pm}\tau^{\mp} \rightarrow \tau^{\pm}\tau^{\mp}\widetilde{\chi}_{1}^{0}$, $\widetilde{\chi}_{2}^{0} \rightarrow h_{0}\widetilde{\chi}_{1}^{0}$, and $\widetilde{\chi}_{2}^{0} \rightarrow Z^{0}\widetilde{\chi}_{1}^{0}$~\cite{Dutta:2008ge}. In the region of the $D$6-brane model parameter space with patterns $\textit{ID6BraneP4}$ and $\textit{ID6BraneP5}$ that can generate the WMAP observed relic density, we see similar states to that of mSUGRA, namely low energy ($\lesssim$20 GeV) opposite sign tau pairs. On the other hand, the states do begin to differ between the $D$6-brane model and mSUGRA in the SSC scenario. We showed that the final states in the SSC region for patterns $\textit{ID6BraneP4}$ and $\textit{ID6BraneP5}$ will be high energy ($\gtrsim$20 GeV) opposite sign tau pairs. High energy tau will be dominant in the region of the mSUGRA parameter space with a large universal gaugino mass $m_{1/2}$, nevertheless, as $m_{1/2}$ is decreased, the dominant decay chains shift to the Higgs boson and Z boson. Therefore, we see similar LHC signals in the SSC region of the $D$6-brane allowed parameter space with patterns $\textit{ID6BraneP4}$ and $\textit{ID6BraneP5}$ and the SSC region of the mSUGRA allowed parameter space only at higher values of $m_{1/2}$. For lower values of $m_{1/2}$ in mSUGRA in the SSC region, there are obvious distinctions with the $D$6-brane model. Clearly identifiable differences exist between the LHC states of mSUGRA and the $D$6-brane model states in the regions of the allowed parameter space with pattern $\textit{ID6BraneP1}$. The decay $\widetilde{\chi}_{2}^{0} \rightarrow \nu\overline{\nu}\widetilde{\chi}_{1}^{0}$ is favored in the $D$6-brane model, but is kinematically forbidden in mSUGRA, and moreover, the production of opposite sign tau pairs is suppressed in the WMAP and SSC regions of the $D$6-brane parameter space with pattern $\textit{ID6BraneP1}$, as compared to mSUGRA.

\section{Neutralino Coannihilation}

\begin{figure}[t]
	\centering
    \includegraphics[width=0.75\textwidth]{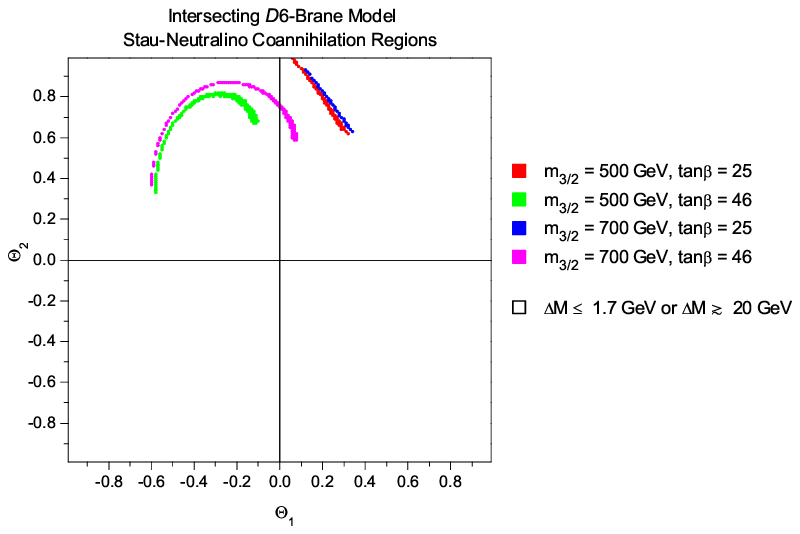}
    \includegraphics[width=0.75\textwidth]{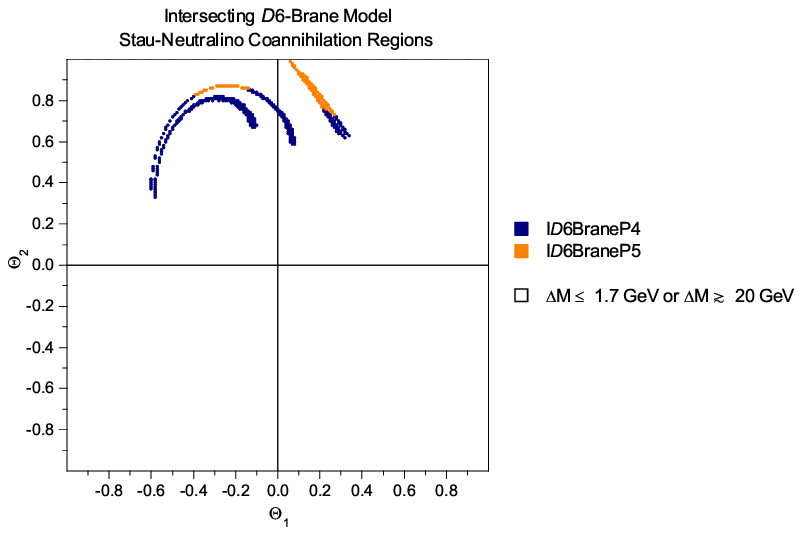}
		\caption{Stau-neutralino coannihilation regions within the intersecting $D$6-Brane model allowed parameter space. The upper plot is differentiated by gravitino mass and tan$\beta$, whilst the lower plot is differentiated by the mass hierarchy patterns. The shaded regions represent 1.7 GeV $<$ $\Delta$M $\lesssim$ 20 GeV, where $\Delta$M = $m_{\widetilde{\tau}_{1}} - m_{\widetilde{\chi}_{1}^{0}}$. These regions will generate the WMAP observed dark matter density.}
	\label{fig:D6_StauCoannihilation}
\end{figure}

We have found that in the region of the allowed parameter space that generates the WMAP constrained dark matter density for patterns $\textit{ID6BraneP4}$ and $\textit{ID6BraneP5}$, the final states of an intersecting $D$6-brane model are essentially the same as the final states for mSUGRA. In mSUGRA, only specific regions of the parameter space are cosmologically allowed within the WMAP dark matter density upper and lower bounds. One of these regions is referred to as the stau-neutralino coannihilation region, where early universe neutralinos can annihilate with stau, producing low-energy tau. It is characterized by a nearly degenerate mass between the lightest neutralino $\widetilde{\chi}_{1}^{0}$ and the tau slepton $\widetilde{\tau}_{1}$, this near degeneracy measured by the mass difference $\Delta$M = $m_{\widetilde{\tau}_{1}} - m_{\widetilde{\chi}_{1}^{0}}$. We found regions of the $D$6-brane model parameter space with stau patterns $\textit{ID6BraneP4}$ and $\textit{ID6BraneP5}$ possess stau-neutralino coannihilation regions, as shown in Fig.~\ref{fig:D6_StauCoannihilation}. The regions plotted in Fig.~\ref{fig:D6_StauCoannihilation} have 1.7 GeV $<$ $\Delta$M $\lesssim$ 20 GeV. In these regions of the $D$6-brane model parameter space, the stau decays to a neutralino and tau 100\% of the time through the process $\widetilde{\tau}_{1}^{\pm}~\rightarrow~\widetilde{\chi}_{1}^{0}\tau^{\pm}$, thus, if $\Delta$M $\leq$ 1.7 GeV, the mass of the tau, then the only evidence of the process will be missing energy. In light of this, for this particular plot we exclude regions of the parameter space where $\Delta$M $\leq$ 1.7 GeV. If we restrict the upper bound to $\sim$20 GeV, then the result will be low energy tau production. Thus, we expect to find similarities between the final states within the shaded regions in Fig.~\ref{fig:D6_StauCoannihilation} and those regions in the coannihilation region of mSUGRA. This will affect how the intersecting $D$6-brane model can be validated at LHC, since any analysis of kinematical variables will have to discriminate between the coannihilation region of mSUGRA and the $\textit{ID6BraneP4}$ and $\textit{ID6BraneP5}$ regions of the $D$6-brane model parameter space. We shall discuss this in more detail shortly.

\begin{figure}[t]
	\centering
    \includegraphics[width=0.75\textwidth]{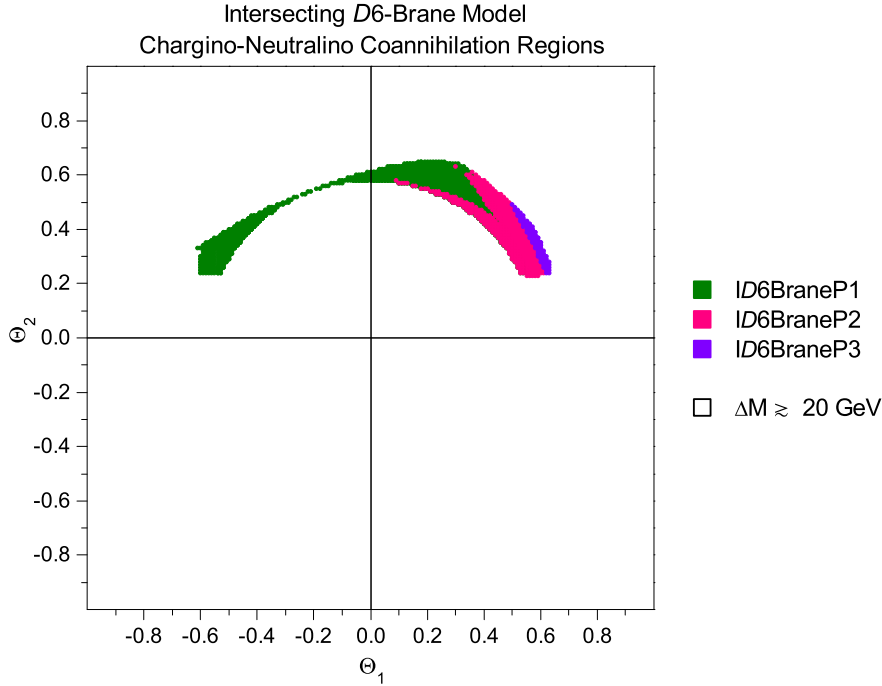}
		\caption{Chargino-neutralino coannihilation regions within the intersecting $D$6-Brane model allowed parameter space. The plot is differentiated by the mass hierarchy patterns. The shaded regions represent a mass difference of $\lesssim$ 20 GeV between the $\widetilde{\chi}_{1}^{\pm}$ or $\widetilde{\chi}_{2}^{0}$ and the $\widetilde{\chi}_{1}^{0}$. Of the three regions plotted here, only the $\textit{ID6BraneP1}$ region will generate the WMAP observed dark matter density and diluted relic density in the SSC scenario. The $\textit{ID6BraneP2}$ and $\textit{ID6BraneP3}$ regions can only generate an extremely small relic density of $\Omega_{\chi}$ $\lesssim$ 0.01.}
	\label{fig:D6_CharginoCoannihilation}
\end{figure}

The regions in Fig.~\ref{fig:D6_ParamSpace} that generate the WMAP observed dark matter density that are not represented in Fig.~\ref{fig:D6_StauCoannihilation} are situated in the chargino-neutralino coannihilation region. In this region in an intersecting $D$6-brane model, the lightest neutralino $\widetilde{\chi}_{1}^{0}$ has a nearly degenerate mass with the lightest chargino $\widetilde{\chi}_{1}^{\pm}$ and second lightest neutralino $\widetilde{\chi}_{2}^{0}$. Here, early universe $\widetilde{\chi}_{1}^{0}$ can annihilate with $\widetilde{\chi}_{1}^{\pm}$ and $\widetilde{\chi}_{2}^{0}$. We found regions of the parameter space with the patterns $\textit{ID6BraneP1}$, $\textit{ID6BraneP2}$, and $\textit{ID6BraneP3}$ containing chargino-neutralino coannihilation regions, as shown in Fig.~\ref{fig:D6_CharginoCoannihilation}. However, only regions of the parameter space with pattern $\textit{ID6BraneP1}$ can generate the WMAP observed relic density in the chargino-coannihilation region. Furthermore, regions of the allowed parameter space with the patterns $\textit{ID6BraneP2}$ and $\textit{ID6BraneP3}$ cannot generate the diluted dark matter density in the SSC scenario either, however, regions with the pattern $\textit{ID6BraneP1}$ can generate the correct SSC relic density. The regions of the parameter space with patterns $\textit{ID6BraneP2}$ and $\textit{ID6BraneP3}$ can only generate an extremely small relic density of $\Omega_{\chi}$ $\lesssim$ 0.01, thus, the neutralino could only comprise a very small portion of the WMAP observed dark matter density. The remainder of the relic density would have to be composed of matter other than neutralinos. We use an upper bound of $\sim$20 GeV in Fig.~\ref{fig:D6_CharginoCoannihilation} to include all of the regions that generate the WMAP relic density. The stau-neutralino coannihilation region in the intersecting $D$6-brane model and in mSUGRA produces low energy tau from stau decays, and this is in contrast to low energy tau in the chargino-neutralino coannihilation region predominantly resulting from $\widetilde{\chi}_{1}^{\pm}$ and $\widetilde{\chi}_{2}^{0}$ decays. This fact will be important when constructing kinematical observables that must distinguish between the intersecting $D$6-brane model and mSUGRA.

\section{Observables and Model Parameter Determination}

The ultimate goal is to determine the model parameters of the intersecting brane model, although this presents new challenges since the soft-terms are in general non-universal. In the present model we have seven soft-supersymmetry breaking mass parameters at the unification scale which are functions of three goldstino angles which parameterize the F-terms, along with the free parameter tan$\beta$. A minimum of four experimental observables are needed to determine the model parameters, where the four observables could be constructed so as to determine four of the eight parameters, say, for example, $M3$, $m_{L}$, $A_{0}$, and tan$\beta$. Once $M3$, $m_{L}$, and $A_{0}$ are determined, then Eqs.~(\ref{eq:idb:gaugino}),~(\ref{eq:idb:trilinear_u}), and~(\ref{eq:idb:scalarmass_u}) can be solved simultaneously for the three free parameters $\Theta_{1}$, $\Theta_{2}$, and $m_{3/2}$. After solving these three equations, these three free parameters can be used to compute the remaining four soft-supersymmetry breaking mass parameters $M2$, $M1$, $m_{R}$, and $m_{H}^{2}$, and henceforth, along with a known tan$\beta$, the sparticle masses and relic density can be computed. It has yet to be determined whether four experimental observables could be constructed to compute four of the eight $D$6-brane model parameters. For mSUGRA, it was shown in~\cite{Arnowitt:2008bz} that the model parameters can be determined for the WMAP constrained region of the relic density, and in~\cite{Dutta:2008ge} in the context of SSC, where four experimental observables were derived, one for each of the four soft-supersymmetry breaking parameters in mSUGRA.

The final states in the regions of the parameter space with pattern $\textit{ID6BraneP1}$ are different than the final states in regions of the parameter space with patterns $\textit{ID6BraneP4}$ and $\textit{ID6BraneP5}$, demonstrated by the fact the $D$6-brane model contains both stau-neutralino and chargino-neutralino coannihilation regions that generate the WMAP observed dark matter density, as well as multiple independent regions that generate the diluted relic density of the SSC scenario. This greatly complicates the task since the construction of experimental observables to determine the model parameters in those regions of the parameter space with pattern $\textit{ID6BraneP1}$ will not necessarily determine the model parameters in those regions of the parameter space with patterns $\textit{ID6BraneP4}$ and $\textit{ID6BraneP5}$. Therefore, with the intersecting $D$6-brane model parameter space as it is currently constrained by Standard Model measurements, it is likely more than four experimental observables will be necessary. The final states in the WMAP and SSC regions of the parameter space are quite similar, though the energy of the lepton pairs will be higher in the SSC region than in the WMAP region. This will necessitate different selection cuts on the data distributions, creating a new experimental observable. Therefore, in this context, the maximum number of observables necessary to determine the model parameters in the $D$6-brane model could well exceed four.

It is essential that an intersecting $D$6-brane model be distinguished from mSUGRA, though this task is complicated by the possibility that the final states of both models are similar in the stau-neutralino coannihilation regions. For the $D$6-brane model, the goal is to build four experimental observables to determine the seven soft-supersymmetry breaking terms and tan$\beta$ by solving for the free parameters and then computing the remaining soft terms. Likewise, it has been shown~\cite{Arnowitt:2008bz}\cite{Dutta:2008ge} that only four observables are necessary to determine the model parameters in mSUGRA, although the universal gaugino and scalar masses in mSUGRA will be different from the non-universal masses in an intersecting brane model. However, since none of the experimentally allowed regions of the $D$6-brane model parameter space that we generated using the equations given in Eqs.~(\ref{eq:idb:gaugino}),~(\ref{eq:idb:bino}),~(\ref{eq:idb:trilinear_u}),~(\ref{eq:idb:scalarmass_u}), and~(\ref{eq:idb:higgsmass_u}) for $\textit{u}$-moduli dominated SUSY breaking have universal gaugino masses and universal scalar masses, mSUGRA is not presumed to be a subset of the $D$6-brane model, and hence, the observables in the $D$6-brane model will most likely possess a different construction than the corresponding observables in mSUGRA.

Construction of experimental observables that can determine model parameters is beyond the scope of this work. In order to do this it is first imperative that the parameter space be further constrained to eventually narrow down the number of different patterns of the mass spectra to only one. This could limit the number of experimental observables necessary to determine the model parameters to four. This can be accomplished by application of new data, both from colliders and from cosmological measurements, such as from direct dark-matter detection and constraints on the galactic gamma flux resulting from neutralino annihilations. For a discussion on direct dark matter detection cross-sections and annihilation rates in the present model, see~\cite{Maxin:2009qq}.

\section{Conclusion}

We have explored the low-energy supersymmetry phenomenology of a near-realistic intersecting $D$6-brane model in Type IIA string theory.  The $D$6 model has three generations of SM fermions and exhibits automatic gauge coupling unification. In addition, it is possible to obtain correct masses and mixings for both up and down-type quarks as well as the tau lepton. To date, this is the only known string model where this is possible.  We calculated the soft-supersymmetry breaking terms and superpartner spectra satisfying all presently known experimental constraints for the $\textit{u}$-moduli dominated SUSY breaking scenario and showed there are regions within the parameter space which may generate both the WMAP observed dark matter density and the diluted relic density in the context of SSC. Five regions in the allowed parameter space were identified that possess a different hierarchy of the four lightest sparticles in the mass spectrum. It was found that only three of these regions can generate the correct WMAP and SSC relic densities. We constructed the final states for regions of the parameter space that can generate the WMAP observed relic density, which consisted of low energy tau and jets in the stau-neutralino coannihilation region, and low energy leptons, high energy neutrinos, and jets in the chargino-neutralino coannihilation region. In the SSC scenario, we found the final states are high energy leptons, high energy neutrinos, and jets. We found that the minimum number of required observables to determine the free parameters is four, although this number of observables could exceed the minimum due to the dissimilar final states between the three regions of the allowed parameter space that can generate the correct relic density. Finally, we discussed how further constraining the parameter space with new measurements can set the maximum number of observables.  Time will tell whether or not string theory will say anything definitive about what is observed at LHC.  

\section{Acknowledgments}

J.M. would like to thank B. Dutta for useful discussions. This research was supported in part by the Mitchell-Heep Chair in High Energy Physics and by DOE grant DE-FG03-95-Er-40917.

\end{document}